\documentclass[twocolumn,twocolappendix]{aastex6}

\usepackage{bm}
\usepackage{amsmath}
\usepackage{xcolor}
\usepackage{graphicx}
\usepackage{hyperref}
\usepackage[all]{hypcap}

\begin{document}

    \title{Propagation of waves in weakly ionized two-fluid plasmas. I. Small-amplitude Alfvénic waves}
    \shorttitle{Alfvénic waves in weakly ionized plasmas}
    \shortauthors{Martínez-Gómez}
    
    \author{David Martínez-Gómez\altaffilmark{1,2}}
    \altaffiltext{1}{Instituto de Astrofísica de Canarias, 38205 La Laguna, Tenerife, Spain}
    \altaffiltext{2}{Departamento de Astrofísica, Universidad de La Laguna, 38205 La Laguna, Tenerife, Spain}
    \email{dmartinez@iac.es}
    
    \begin{abstract}
        The large abundance of electrically neutral particles has a remarkable impact on the dynamics of many astrophysical plasmas. Here, we use a two-fluid model that includes charge-neutral elastic collisions and Hall's current to study the propagation of magnetohydrodynamic (MHD) waves in weakly ionized plasmas. We derive the dispersion relation for small-amplitude incompressible transverse waves propagating along the background magnetic field. Then, we focus on the polarization relations fulfilled by the eigenmodes and their corresponding ratios of magnetic to kinetic energies, and we study their dependence on the relations between the oscillation, collision and cyclotron frequencies. For low wave frequencies, the two components of the plasma are strongly coupled, the damping due to the charge-neutral interaction is weak and the effect of Hall's term is negligible. However, as the wave frequency increases, phase shifts between the velocity of charges, the velocity of neutrals, and the magnetic field appear, leading to enhanced damping. The effect of collisions on the propagation of waves strongly depends on their polarization state, with the left-handed circularly polarized ion-cyclotron modes being more efficiently damped than the linearly polarized Alfvén waves and the right-handed circularly polarized whistler modes. Moreover, the equipartition relation between the magnetic energy and the kinetic energy of Alfvén waves does not hold in general when the collisional interaction and Hall's current are taken into account, with the magnetic energy usually dominating over the kinetic energy. This theoretical result extends previous findings from observational and numerical works about turbulence in astrophysical scenarios.
    \end{abstract}
    

\section{Introduction} \label{sec:intro}
    Plasmas in many astrophysical environments are not fully ionized but partially ionized, meaning that they contain a non-negligible fraction of electrically neutral particles. This is the case, for instance, of molecular clouds, the interstellar medium, planetary ionospheres and the lower layers of the solar atmosphere \citep[see, e.g.,][]{Ballester2018SSRv..214...58B}. The presence of this neutral component, which is not directly affected by magnetic fields, may have a strong impact on the dynamics of the plasma \citep[see, e.g.,][]{Mestel1956MNRAS.116..503M,Piddington1956MNRAS.116..314P,Watanabe1961CaJPh..39.1044W,Kulsrud1969ApJ...156..445K}. For instance, the momentum exchange between the electrically charged and neutral particles by means of collisions leads to the damping of magnetohydrodynamic (MHD) waves \citep{Piddington1956MNRAS.116..314P,Watanabe1961CaJPh..39.1197W} but it also may cause a reduction in their propagation speeds \citep{Kumar2003SoPh..214..241K,Soler2013ApJS..209...16S} and modify the structure of shocks \citep{Hillier2016A&A...591A.112H,Snow2021A&A...645A..81S}. In addition, the dissipation of the wave energy due to the collisional interaction may generate an important heating of the plasma \citep{Leake2005A&A...442.1091L,Song2011JGRA..116.9104S,KhomenkoCollados2012ApJ...747...87K,MartinezSykora2012ApJ...753..161M,Arber2016ApJ...817...94A}.

    When the oscillation frequencies of the waves are much smaller than all the frequencies associated with the collisional processes, the different components of the plasma are strongly coupled and their interaction can be accurately described through single-fluid models. This approach has been used by, for instance, \citet{DePontieu2001ApJ...558..859D}, \citet{Khodachenko2004A&A...422.1073K}, \citet{Forteza2007A&A...461..731F,Forteza2008A&A...492..223F} and \citet{Soler2010A&A...512A..28S}, to study the properties of the three classic types of MHD waves, that is, the incompressible Alfvén modes and the compressible fast and slow magnetoacoustic modes. These works have found that charge-neutral collisions are typically more efficient in dissipating the magnetic energy than the acoustic energy of the waves \citep{Soler2015ApJ...810..146S} and that the efficiency of energy dissipation increases with the wave frequency \citep{Shelyag2016ApJ...819L..11S}.

    On the other hand, if the wave frequencies are comparable or larger than the collision frequencies, the coupling between the various species of the plasma becomes weaker and large drift velocities may appear, as it has been shown by means of observations \citep[see, e.g.,][]{Khomenko2016ApJ...823..132K,Anan2017A&A...601A.103A,Stellmacher2017SoPh..292...83S,Wiehr2019ApJ...873..125W,Wiehr2021ApJ...920...47W,Zapior2022ApJ...934...16Z,Gonzalez2024A&A...681A.114G} and numerical simulations \citep[see, e.g.,][]{Hillier2019PhPl...26h2902H,Popescu2021A&A...646A..93P,Popescu2021A&A...650A.181P,MartinezGomez2022ApJ...940L..47M}. In these conditions, the dynamics of the plasma is better described by multi-fluid models, which have been commonly applied in investigations of waves and hydrodynamic instabilities in molecular clouds \citep{Pudritz1990ApJ...350..195P,Balsara1996ApJ...465..775B,Mouschovias2011MNRAS.415.1751M}, planetary ionospheres \citep{Schunk1971JGR....76.6159S,Leake2014SSRv..184..107L}, the interstellar medium \citep{Pinto2008A&A...484...17P} or the solar atmosphere \citep{Zaqarashvili2011A&A...529A..82Z,Khomenko2014PhPl...21i2901K,Soler2016A&A...592A..28S,Soler2017ApJ...840...20S}. Multi-fluid models have a larger range of applicability than single-fluid methods and reveal effects that cannot be fully captured by the latter. For instance, using a two-fluid model that separates the charged species from the neutral ones, \citet{Kulsrud1969ApJ...156..445K}, \citet{Kamaya1998ApJ...500..257K}, and \citet{Soler2013ApJ...767..171S,Soler2013ApJS..209...16S} showed the existence of intervals of wavelengths in which there are no oscillatory standing MHD waves in weakly ionized plasmas: in these so-called cutoff regions the friction force due to collisions between ions and neutrals dominates over the restoring force of the magnetic tension. Moreover, the numerical studies by \citet{Leake2012ApJ...760..109L,Hillier2016A&A...591A.112H,Murtas2021PhPl...28c2901M,Murtas2022PhPl...29f2302M} have shown that the decoupling between ions and neutrals and the processes of ionization and recombination play an important role in the phenomenon of magnetic reconnection.

    The present paper is framed in the context of two-fluid modeling. In particular, this is the first installment of a series with the goal of studying the properties of MHD waves propagating in weakly ionized plasmas. Here, we focus on the case of small-amplitude incompressible waves while the study of the effects of compressibility and non-linearities is left for future works. Following a similar procedure to that employed by \citet{Zaqarashvili2011A&A...529A..82Z} and \citet{Soler2013ApJ...767..171S}, we first consider the linear regime of the two-fluid equations to derive the dispersion relation for small-amplitude Alfvén waves and analyze how their wavenumbers, damping rates, quality factors, and phase speeds depend on the relation between the wave frequency and the collision frequencies. Then, we complement these results with the study of the eigenfunction relations (or polarization relations) fulfilled by the eigenmodes (or normal modes) of the two-fluid system, which is a matter that has previously received little to no attention (one of the few examples of the use of this procedure can be found in \citet{Ofman2005JGRA..110.9102O}, who applied it to the case of multi-ion plasmas in the solar corona). This analysis allows us to check the existence of phase shifts between the different variables of relevance (velocity of charges, velocity of neutrals, and magnetic field) and how these phase shifts depend on the coupling degree between the two fluids.

    Next, we explore the impact of taking into account Hall's term in the induction equation. It is known that in fully ionized plasmas, the effect of this term only becomes of special relevance when the wave frequency approaches the frequencies associated to the cyclotron motions of the ions \citep{Lighthill1960RSPTA.252..397L,Stix1992wapl.book.....S,Cramer2001paw..book.....C}, strongly modifying the properties of waves. For instance, the linearly polarized Alfvén waves split into two different circularly polarized modes: the left-handed ion-cyclotron and the right-handed whistler modes (assuming that the wave frequency is positive). However, it has also been shown that in weakly ionized media the collisional coupling of the charges with the neutral component of the plasma produces a reduced effective cyclotron frequency and the influence of Hall's term becomes important at much lower frequencies and smaller spatial scales \citep{Amagishi1993PhRvL..71..360A,Pandey2008MNRAS.385.2269P,Pandey2015MNRAS.447.3604P}. Since we are interested in studying a wide range of frequencies, it is important to consider the combined effects of elastic collisions and Hall's current and to analyze how the properties of the waves vary with their polarization states.

    Finally, relying on the results from the investigation described in the previous paragraphs, we move a step further and study the energy density of the waves and how it is distributed between its kinetic and magnetic components. In this way, we analyze how the Walén relation, which states that in fully ionized plasmas there is equipartition between the magnetic and kinetic energies of Alfvén waves \citep{Walen1944ArMAF..30A...1W,Ferraro1958ApJ...127..459F,Braginskii1965RvPP....1..205B,Priest1984smh..book.....P}, varies in a more general scenario. It has already been found in observations of the interplanetary plasma \citep{Belcher1971JGR....76.3534B,Matthaeus1982JGR....87.6011M,Bruno1985JGR....90.4373B} and numerical simulations of turbulence in the solar wind \citep{Belcher1969JGR....74.2302B,Tanenbaum1962PhFl....5.1226T,Dastgeer2000PhPl....7..571D,Boldyrev2012AIPC.1436...18B,Wang2011ApJ...740L..36W} that the equipartition relation is not always fulfilled, with the magnetic energy being usually larger than the kinetic energy. This finding has been explained in terms of the effect of Hall's current at short scales \citep{Campos1992_10.1063/1.860136,Galtier2006JPlPh..72..721G,Lingam2016MNRAS.460..478L,Lingam2016ApJ...829...51L}. In addition, in their study about turbulence in partially ionized molecular clouds, \citet{Zweibel1995ApJ...439..779Z} found that wave damping due to ion-neutral collisions has a small effect on the equipartition of energy. However, these authors did not take into account Hall's term and assumed a strong collisional coupling which results in a weak damping of the waves. Here, we expand this research to a broader range of the collisional coupling and explore whether the charge-neutral interaction, combined with the effect of Hall's current, may have a large impact on the ratio between the magnetic and kinetic energies for the three different kind of waves considered in this work (Alfvén, ion-cyclotron and whistler modes).

    The outline of this paper is as follows. In Section \ref{sec:model} we detail the equations of the two-fluid model and the background atmosphere that we use as reference to represent a stratified weakly ionized medium. Section \ref{sec:methods} shows the derivation of the dispersion relation for small-amplitude Alfvénic waves. In Section \ref{sec:results} we analyze the properties of the normal modes, focusing on the influence of the charge-neutral collisions and Hall's current. Finally, Section \ref{sec:concl} contains a summary of the main results and a discussion of their implications for both observational and theoretical investigations of partially ionized plasmas.
    
\section{Model} \label{sec:model}
\subsection{Two-fluid equations}
    In this work we study the dynamics of partially ionized plasmas by means of a two-fluid approach \cite[see, e.g.,][]{Mouschovias2011MNRAS.415.1751M,Zaqarashvili2011A&A...529A..82Z,Khomenko2014PhPl...21i2901K,Popescu2019A&A...627A..25P}. We assume that there is a strong coupling between ions (i) and electrons (e), so they can be treated together as one charged fluid (c), while the other fluid contains the neutral component of the plasma (n). Both fluids are allowed to interact by means of elastic collisions only, so there is no ionization or recombination. In addition, we take into account the effect of Hall's current \citep{Lighthill1960RSPTA.252..397L,Cramer2001paw..book.....C} but neglect other non-ideal processes such as Ohmic diffusion, viscosity, thermal conduction or anisotropy of temperature. Therefore, the non-linear equations that describe the temporal evolution of the plasma are given by
    \begin{equation} \label{eq:rho_n}
        \frac{\partial \rho_{\rm{n}}}{\partial t} + \nabla \cdot \left(\rho_{\rm{n}} \bm{V}_{\rm{n}} \right) = 0,
    \end{equation}

    \begin{equation} \label{eq:rho_c}
        \frac{\partial \rho_{\rm{c}}}{\partial t} + \nabla \cdot \left(\rho_{\rm{c}} \bm{V}_{\rm{c}} \right) = 0,
    \end{equation}

    \begin{equation} \label{eq:mom_n}
        \frac{\partial \left(\rho_{\rm{n}} \bm{V}_{\rm{n}} \right)}{\partial t} + \nabla \cdot \left(\rho_{\rm{n}} \bm{V}_{\rm{n}} \bm{V}_{\rm{n}} \right) = - \nabla P_{\rm{n}} + \rho_{\rm{n}} \bm{g} + \bm{R}_{\rm{n}},
    \end{equation}

    \begin{equation} \label{eq:mom_c}
        \frac{\partial \left(\rho_{\rm{c}} \bm{V}_{\rm{c}} \right)}{\partial t} + \nabla \cdot \left(\rho_{\rm{c}} \bm{V}_{\rm{c}} \bm{V}_{\rm{c}} \right) = - \nabla P_{\rm{c}} + \frac{\nabla \times \bm{B}}{\mu_{0}} \times \bm{B} + \rho_{\rm{c}} \bm{g} + \bm{R}_{\rm{c}},
    \end{equation}

    \begin{equation} \label{eq:pres_n}
        \frac{\partial P_{\rm{n}}}{\partial t} +\left(\bm{V}_{\rm{n}} \cdot \nabla \right) P_{\rm{n}} + \gamma P_{\rm{n}} \nabla \cdot \bm{V}_{\rm{n}} = \left(\gamma - 1\right) Q_{\rm{nc}},
    \end{equation}

    \begin{equation} \label{eq:pres_c}
        \frac{\partial P_{\rm{c}}}{\partial t} +\left(\bm{V}_{\rm{c}} \cdot \nabla \right) P_{\rm{c}} + \gamma P_{\rm{c}} \nabla \cdot \bm{V}_{\rm{c}} = \left(\gamma - 1\right) Q_{\rm{cn}},
    \end{equation}

    \begin{equation} \label{eq:indu}
        \frac{\partial \bm{B}}{\partial t} = \nabla \times \left(\bm{V}_{\rm{c}} \times \bm{B}\right) - \nabla \times \left[\frac{\left(\nabla \times \bm{B} \right) \times \bm{B}}{e n_{\rm{e}} \mu_{0}}\right], 
    \end{equation}
    where $\rho_{\rm{n}}$ and $\rho_{\rm{c}}$ are the densities of the neutral and charged fluid, $\bm{V}_{\rm{n}}$ and $\bm{V}_{\rm{c}}$ their respective velocities, $P_{\rm{n}}$ and $P_{\rm{c}}$ the pressures, $\bm{B}$ the magnetic field, and $n_{\rm{e}}$ the number density of electrons. The density and pressure of the charged fluid contain the contributions from ions and electrons, so $\rho_{\rm{c}} = \rho_{\rm{i}} + \rho_{\rm{e}}$ and $P_{\rm{c}} = P_{\rm{i}} + P_{\rm{e}}$. The acceleration of gravity is denoted by $\bm{g}$, the vacuum magnetic permeability by $\mu_{0}$, the adiabatic constant by $\gamma$, and the elementary electric charge by $e$.

    The terms $\bm{R}_{\rm{n}}$, $\bm{R}_{\rm{c}}$, $Q_{\rm{cn}}$, and $Q_{\rm{nc}}$ describe the momentum and energy transfer due to elastic collisions between the two fluids \citep{Schunk1977RvGSP..15..429S,Draine1986MNRAS.220..133D} and are given by 
    \begin{equation} \label{eq:rn_rc}
        \bm{R}_{\rm{n}} = \alpha_{\rm{nc}} \left(\bm{V}_{\rm{c}} - \bm{V}_{\rm{n}} \right), \quad \bm{R}_{\rm{c}} = \alpha_{\rm{cn}} \left(\bm{V}_{\rm{n}} - \bm{V}_{\rm{c}} \right),        
    \end{equation}
    \begin{equation} \label{eq:qn}
        Q_{\rm{nc}} = \frac{2 \alpha_{\rm{nc}}}{m_{\rm{n}} + m_{\rm{c}}} \left[\frac{3}{2}k_{\rm{B}} \left(T_{\rm{c}} - T_{\rm{n}}\right) + \frac{1}{2}m_{\rm{c}} \left(\bm{V}_{\rm{c}} - \bm{V}_{\rm{n}}\right)^{2} \right],
    \end{equation}

    \begin{equation} \label{eq:qc}
        Q_{\rm{cn}} = \frac{2 \alpha_{\rm{cn}}}{m_{\rm{c}} + m_{\rm{n}}} \left[\frac{3}{2}k_{\rm{B}} \left(T_{\rm{n}} - T_{\rm{c}}\right) + \frac{1}{2}m_{\rm{n}} \left(\bm{V}_{\rm{n}} - \bm{V}_{\rm{c}}\right)^{2} \right],
    \end{equation}
    where $\alpha_{\rm{nc}}$ and $\alpha_{\rm{cn}}$ are the collisional friction coefficients, $m_{\rm{c}}$ and $m_{\rm{n}}$ are the masses of charges and neutrals, respectively, $k_{\rm{B}}$ is the Boltzmann constant, and $T_{\rm{n}}$ and $T_{\rm{c}}$ are the temperatures of the fluids. As noted by \citet{Schunk1977RvGSP..15..429S}, in general the momentum and energy transfer terms should be multiplied by some correction factors that depend on the ratio between the drift speed, $|\bm{V}_{\rm{n}} - \bm{V}_{\rm{c}}|$, and the reduced thermal speed, $V_{\rm{therm}} \equiv \sqrt{2 k_{\rm{B}} \left(m_{\rm{c}} T_{\rm{n}} + m_{\rm{n}}T_{\rm{c}}\right)/\left(m_{\rm{c}}m_{\rm{n}}\right)}$. However, in the limit of small relative drift speeds, which is valid for the present investigation, it is a good approximation to not include those correction factors.

    Due to the conservation of momentum, the relations $\bm{R}_{\rm{n}} = - \bm{R}_{\rm{c}}$ and $\alpha_{\rm{nc}} = \alpha_{\rm{cn}}$ are fulfilled. However, it must be noted that $Q_{\rm{nc}} \ne Q_{\rm{cn}}$. The friction coefficient for collisions between charged and neutral species is given by \citep{Braginskii1965RvPP....1..205B,Draine1986MNRAS.220..133D}
    \begin{equation} \label{eq:alpha_cn}
        \alpha_{\rm{cn}} = \frac{\rho_{\rm{c}}\rho_{\rm{n}}}{m_{\rm{c}}+m_{\rm{n}}} \sqrt{\frac{8 k_{\rm{B}}}{\pi}\left(\frac{T_{\rm{c}}}{m_{\rm{c}}} + \frac{T_{\rm{n}}}{m_{\rm{n}}}\right)} \sigma_{\rm{cn}},
    \end{equation}
    where $\sigma_{\rm{cn}}$ is the collisional cross section. The neutral-charge and the charge-neutral collision frequencies ($\nu_{\rm{nc}}$ and $\nu_{\rm{cn}}$, respectively) are related to the friction coefficient by $\alpha_{\rm{cn}} = \rho_{\rm{n}} \nu_{\rm{nc}} = \rho_{\rm{c}} \nu_{\rm{cn}}$.

    Finally, we assume that both fluids follow the ideal equation of state, so
    \begin{equation} \label{eq:state}
        P_{\rm{n}} = n_{\rm{n}} k_{\rm{B}} T_{\rm{n}}, \quad P_{\rm{c}} = n_{\rm{c}} k_{\rm{B}} T_{\rm{c}},
    \end{equation}
    where $n_{\rm{n}}$ and $n_{c}$ are the number densities of neutrals and charges. The latter contains the contributions from ions and electrons: $n_{\rm{c}} = n_{\rm{i}} + n_{\rm{e}}$.

\subsection{Background atmosphere} \label{sec:background}
    In the present investigation we use the solar chromosphere as an example of a weakly ionized environment. For the sake of simplicity, we represent the chromosphere as an isothermal vertically stratified medium embedded in a uniform magnetic field, which is a modified version of the atmospheric model used by \citet{Popescu2019A&A...627A..25P,Popescu2019A&A...630A..79P}. We also assume that the plasma is only composed of hydrogen, which is a valid assumption taking into account its large abundance in comparison with the rest of elements in this layer of the solar atmosphere \citep[see, e.g.,][]{Fontenla1993ApJ...406..319F}. Therefore, we have that $m_{\rm{n}} = m_{\rm{c}} = m_{\rm{p}}$, where $m_{\rm{p}}$ is the proton mass, and that $n_{\rm{i}} = n_{\rm{e}}$, so $n_{\rm{c}} = 2 n_{\rm{e}}$. Densities and number densities are thus related by $\rho_{\rm{n}} = m_{\rm{p}} n_{\rm{n}}$ and $\rho_{\rm{c}} = m_{\rm{p}} n_{\rm{c}}/2$.

    Assuming that $\bm{g} = \left(0, 0, -g \right)$ and that the background magnetic field is given by $\bm{B}_{0} = \left(0, 0, B_{0} \right)$, the conditions of magneto-hydrostatic equilibrium derived from Eqs. (\ref{eq:mom_n}) and (\ref{eq:mom_c}) lead to the following expressions for the background densities and pressures as functions of height, $z$:
    \begin{equation} \label{eq:equil_rho}
        \rho_{\rm{n0}}(z) = \rho_{\rm{n,bot}} e^{-z / H_{\rm{n}}}, \quad \rho_{\rm{c0}}(z) = \rho_{\rm{c,bot}} e^{-z / H_{\rm{c}}},
    \end{equation}
    \begin{equation} \label{eq:equil_pres}
        P_{\rm{n0}}(z) = P_{\rm{n,bot}} e^{-z / H_{\rm{n}}}, \quad
        P_{\rm{c0}}(z) = P_{\rm{c,bot}} e^{-z / H_{\rm{c}}},
    \end{equation}
    where $\rho_{\rm{n,bot}}$, $\rho_{\rm{c,bot}}$, $P_{\rm{n,bot}}$, and $P_{\rm{c,bot}}$ represent the values at $z = 0$. The parameters $H_{\rm{n}}$ and $H_{\rm{c}}$ are the vertical scale heights, given by
    \begin{equation} \label{eq:hn_hc}
        H_{\rm{n}} = \frac{k_{\rm{B}}T_{\rm{n0}}}{m_{\rm{p}}g}, \quad \text{and} \quad H_{\rm{c}} = \frac{2k_{\rm{B}}T_{\rm{c0}}}{m_{\rm{p}}g},
    \end{equation}
    with $T_{\rm{n0}}$ and $T_{\rm{c0}}$ the initial temperatures of the fluids.

    The total density of the plasma as a function of height is then $\rho_{\rm{0}}(z) = \rho_{\rm{n0}}(z) + \rho_{\rm{c0}}(z)$, and the relative densities of each component can be defined as $\xi_{\rm{n0}}(z) = \rho_{\rm{n0}}(z) / \rho_{\rm{0}}(z)$ and $\xi_{\rm{c0}}(z) = \rho_{\rm{c0}}(z)/ \rho_{\rm{0}}(z)$, which fulfill that $\xi_{\rm{c0}}(z) + \xi_{\rm{n0}}(z) = 1$. As shown by \citet{Soler2013ApJ...767..171S}, it is also useful to define the neutral-to-charge density ratio as
    \begin{equation} \label{eq:ionization_ratio}
        \chi (z) = \frac{\rho_{\rm{n0}}(z)}{\rho_{\rm{c0}}(z)},
    \end{equation}
    since this factor will appear in many of the formulas that describe the properties of the magnetohydrodynamic waves.

\begin{figure}
        \includegraphics[width=\hsize]{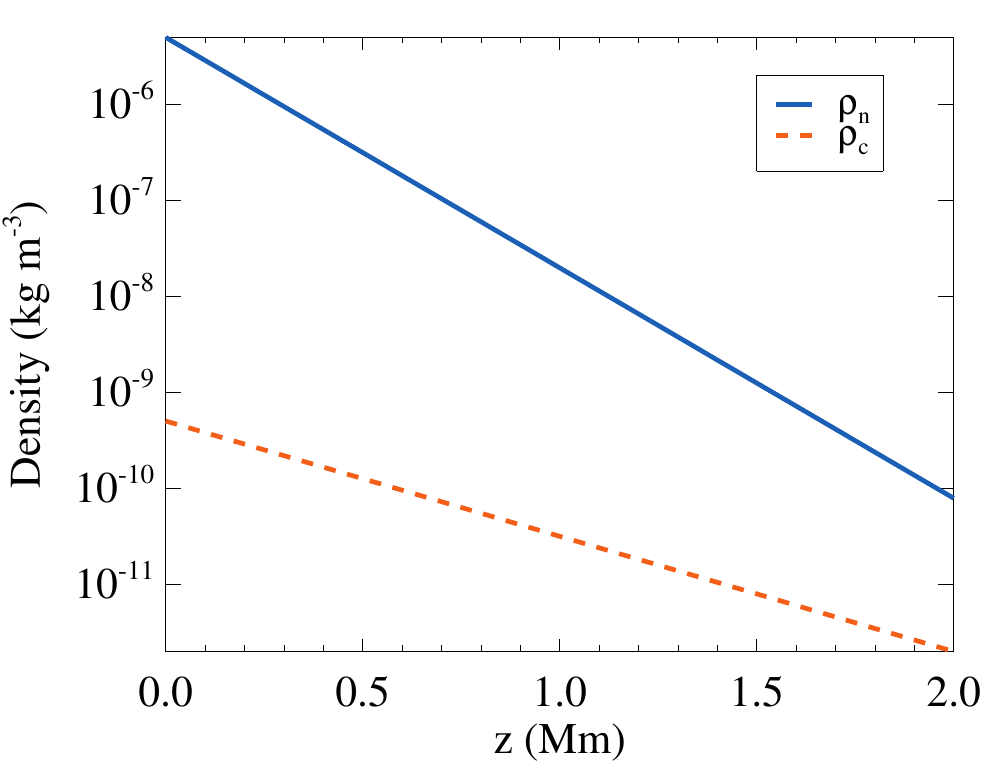}
        \caption{Density of neutral fluid (blue solid line) and charged fluid (dashed orange line) as functions of height.}
        \label{fig:rhos}
    \end{figure}

    For this study, we choose the following values of the atmospheric parameters: $\rho_{\rm{0}}(z=0) = 5 \times 10^{-6} \ \rm{kg \ m^{-3}}$, $\xi_{\rm{c0}}(z=0) = 10^{-4}$, $g = 273.98 \ \rm{m \ s^{-2}}$, $B_{\rm{0}} = 10 \ \rm{G}$, and $T_{\rm{c0}} = T_{\rm{n0}} = 6000 \ \rm{K}$. Consequently, the vertical scale heights are $H_{\rm{n}} \approx 181 \ \rm{km}$ and $H_{\rm{c}} \approx 362 \ \rm{km}$. The density profiles of charges and neutrals as functions of height are represented in Fig. \ref{fig:rhos}, where it can be seen that $\rho_{\rm{n0}}(z) \gg \rho_{\rm{c0}}(z)$ in the whole atmosphere but the neutral-to-charge density ratio decreases with height, going from a value of $\chi (z) = \rho_{\rm{n0}}(z)/\rho_{\rm{c0}}(z) \approx 10^{4}$ at the bottom of the atmosphere to a value of $\chi (z) \approx 40$ at the top.

    \begin{figure}
        \includegraphics[width=\hsize]{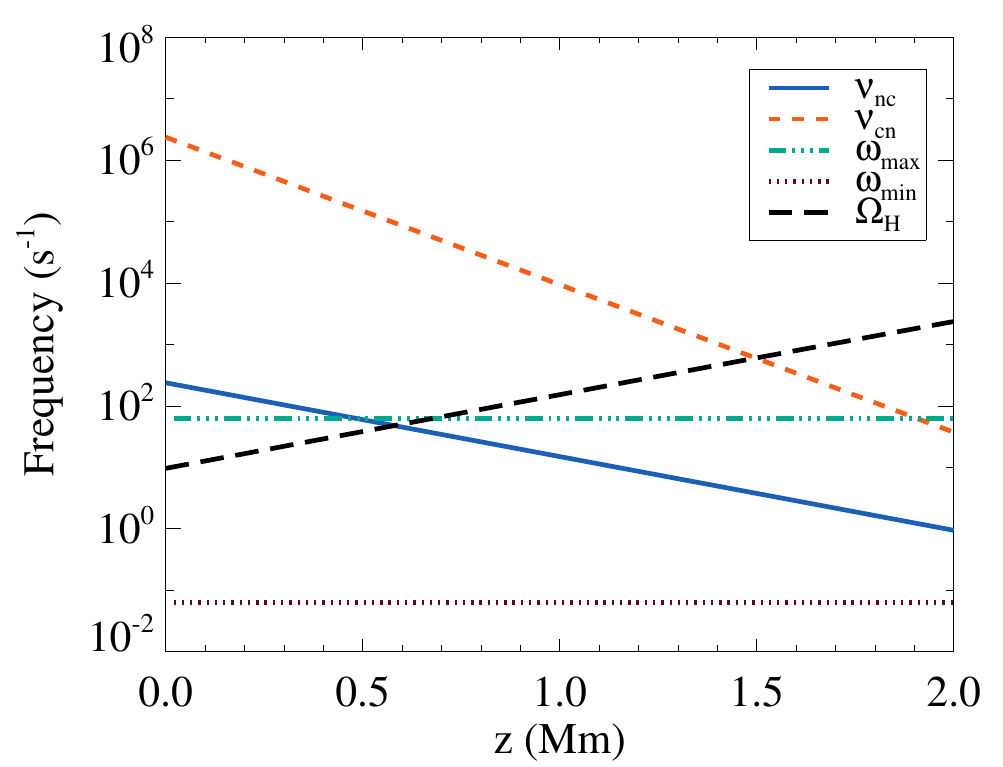}
        \caption{Dependence on height of the collisional frequencies $\nu_{\rm{nc}}$ and $\nu_{\rm{cn}}$, and the Hall frequency, $\Omega_{\rm{H}}$. For reference, the maximum and minimum angular frequencies ($\omega_{\rm{max}}$ and $\omega_{\rm{min}}$, respectively) of the waves considered in the study are shown as horizontal lines.}
        \label{fig:freqs}
    \end{figure}
    
    Then, Fig. \ref{fig:freqs} shows the dependence on height of the relevant frequencies for the present investigation. The collisional frequencies $\nu_{\rm{nc}}$ and $\nu_{\rm{cn}}$ have been computed using a collisional cross section of $\sigma_{\rm{cn}} = 10^{-19} \ \rm{m^{2}}$ \citep{Draine1980ApJ...241.1021D,Draine1986MNRAS.220..133D}. Discussions on the effect that the precise value of the cross-section has on the dynamics of partially ionized plasmas can be found, for instance, in \citet{Pinto2008A&A...484...17P}, \citet{Vranjes2013A&A...554A..22V}, \citet{Soler2015A&A...573A..79S} or \citet{Wargnier2022ApJ...933..205W}. We note here that we have dropped the subscript ``0'' from the symbols for the collision frequencies because in the remaining of the paper we will assume that these parameters do not vary with time. 
    
    The black long-dashed line in Fig. \ref{fig:freqs} represents the so-called Hall frequency \citep{Amagishi1993PhRvL..71..360A,Pandey2008MNRAS.385.2269P}, defined as
    \begin{equation} \label{eq:omega_hall}
        \Omega_{\rm{H}} = \frac{\Omega_{\rm{i}}}{1 + \chi},
    \end{equation}
    where $\Omega_{\rm{i}} = e B_{\rm{0}}/m_{\rm{i}}$ is the cyclotron frequency of ions (here $m_{\rm{i}} = m_{\rm{p}}$). In a partially ionized plasma, the Hall frequency may play an analogous role to that of the ion cyclotron frequency in fully ionized plasmas, as it has been shown by \citet{Amagishi1993PhRvL..71..360A}, \citet{Pandey2008MNRAS.385.2269P}, \citet{Pandey2015MNRAS.447.3604P} or \citet{MartinezGomez2017ApJ...837...80M}. The influence of this parameter on the propagation of Alfvénic waves will be explored in the following sections. 

    Finally, the horizontal lines included in Fig. \ref{fig:freqs} depict the range of wave frequencies that we will consider in this paper. We will study the case of 4 periods of different orders of magnitude. The chosen values are $\tau_{\rm{1}} = 0.1 \ \rm{s}$, $\tau_{\rm{2}} = 1 \ \rm{s}$, $\tau_{\rm{3}} = 10 \ \rm{s}$, and $\tau_{\rm{4}} = 100 \ \rm{s}$, which correspond to the angular frequencies $\omega_{\rm{max}} \equiv \omega_{\rm{1}} = 20 \pi \ \rm{rad \ s^{-1}}$, $\omega_{\rm{2}} = 2\pi \ \rm{rad \ s^{-1}}$, $\omega_{\rm{3}} = 0.2 \pi \ \rm{rad \ s^{-1}}$, and $\omega_{\rm{min}} \equiv \omega_{\rm{4}} = 0.02 \pi \ \rm{rad \ s^{-1}}$. With this selection we cover a wide range of scenarios in relation to the comparison with respect to the characteristic frequencies of the plasma, which will have a strong influence on the behavior of the MHD waves, as it will be shown later. For instance, we see in Fig. \ref{fig:freqs} that $\omega_{\rm{min}} \ll \nu_{\rm{nc}}, \nu_{\rm{cn}}, \Omega_{\rm{H}}$ in the whole atmosphere, corresponding to a case of strong collisional coupling between the two fluids and where the Hall term is not expected to play a relevant role. However, as the frequency of the waves is increased, we may find regions of the atmosphere where $\nu_{\rm{nc}} \lesssim \omega \ll \nu_{\rm{cn}}$ and the collisional coupling becomes weaker. In addition, for the largest wave frequency we have three main different situations: 1) $\Omega_{\rm{H}} \lesssim \omega_{\rm{max}} \lesssim \nu_{\rm{nc}}$ and $\omega_{\rm{max}} \ll \nu_{\rm{cn}}$ at the bottom of the atmosphere (here the Hall term is expected to be of large relevance); 2) $\nu_{\rm{nc}} \lesssim \omega_{\rm{max}} \lesssim \Omega_{\rm{H}}$ at the middle heights; and 3) $\\nu_{\rm{nc}} \ll \omega_{\rm{max}} \sim \nu_{\rm{cn}}$ at the top of the atmosphere, so a weak coupling is expected in this region.  

\section{Methods} \label{sec:methods}
\subsection{Linear regime} \label{sec:linear}
    Since we are interested in the properties of small amplitude waves, we can reduce the system of equations (\ref{eq:rho_n})-(\ref{eq:indu}) to its linearized version. To do so, we assume that each variable (except velocities) can be expressed as the combination of its respective background value (denoted by the subscript ``0'') and a small perturbation (denoted by the subscript ``1''), as $\bm{f} = \bm{f}_{\rm{0}} + \bm{f}_{\rm{1}}$. For the case of velocities and since we consider a static background, we have that $\bm{V}_{\rm{c0}} = \bm{V}_{\rm{n0}} = 0$, so $\bm{V}_{\rm{c}} = \bm{V}_{\rm{c1}}$ and $\bm{V}_{\rm{n}} = \bm{V}_{\rm{n1}}$. The amplitude of the velocity perturbations is considered small in comparison with the characteristic speeds of the system, which in the present case will be related to the Alfvén speed (the definition of this parameter will be shown later). After applying these assumptions, we obtain the following set of linearized equations:
    \begin{equation} \label{eq:lin_rho_n}
        \frac{\partial \rho_{\rm{n}1}}{\partial t} +\rho_{\rm{n}0} \nabla \cdot \bm{V}_{\rm{n}1} = 0, 
    \end{equation}
    
    \begin{equation} \label{eq:lin_rho_c}
        \frac{\partial \rho_{\rm{c}1}}{\partial t} +\rho_{\rm{c}0} \nabla \cdot \bm{V}_{\rm{c}1} = 0,
    \end{equation}
  
    \begin{equation} \label{eq:lin_momn}
        \rho_{\rm{n}0} \frac{\partial \bm{V}_{\rm{n}1}}{\partial t} = -\nabla P_{\rm{n}1} + \alpha_{\rm{cn}} \left(\bm{V}_{\rm{c}1}  - \bm{V}_{\rm{n}1}\right),
    \end{equation}
   
    \begin{equation} \label{eq:lin_momc}
        \rho_{\rm{c}0} \frac{\partial \bm{V}_{\rm{c}1}}{\partial t} = -\nabla P_{\rm{c}1} + \frac{\nabla \times \bm{B}_{\rm{1}}}{\mu_{\rm{0}}} \times \bm{B}_{\rm{0}} 
        - \alpha_{\rm{cn}} \left(\bm{V}_{\rm{c}1} - \bm{V}_{\rm{n}1}\right),
    \end{equation}
   
    \begin{equation} \label{eq:lin_presn}
        \frac{\partial P_{\rm{n}1}}{\partial t} + \gamma P_{\rm{n}0} \nabla \cdot \bm{V}_{\rm{n}1} = 0,
    \end{equation}
    
    \begin{equation} \label{eq:lin_presc}
        \frac{\partial P_{\rm{c}1}}{\partial t} + \gamma P_{\rm{c}0} \nabla \cdot \bm{V}_{\rm{c}1} = 0,
    \end{equation}
    
    \begin{equation} \label{eq:lin_induc}
        \frac{\partial \bm{B}_{1}}{\partial t} = \nabla \times \left(\bm{V}_{\rm{c1}} \times \bm{B}_{\rm{0}} \right) - \nabla \times \left[\frac{\left(\nabla \times \bm{B}_{\rm{1}} \right) \times \bm{B}_{\rm{0}}}{e n_{\rm{e0}} \mu_{\rm{0}}} \right]
    \end{equation}

    In the equations above we have retained the Hall term in the induction equation, that is the last term in Eq. (\ref{eq:lin_induc}), but we have neglected the gravity terms in the momentum equations of both fluids, although we are interested in the propagation of waves in a gravitationally stratified atmosphere. We justify this choice by assuming that the wavelength of the waves of interest is much shorter than the vertical scale heights due to gravity, $H_{\rm{c}}$ and $H_{\rm{n}}$, so the gravitational stratification does not affect the wavelengths or damping rates due to the collisional interaction. Therefore, the system of equations (\ref{eq:lin_rho_n})-(\ref{eq:lin_induc}) can be independently applied to each height of the atmospheric model represented in Fig. \ref{fig:rhos}, in a similar way as it has been done in, for instance, \citet{DePontieu1998A&A...338..729D} and \citet{Soler2013ApJ...767..171S,Soler2013ApJS..209...16S}. It is known, nonetheless, that the gravitational stratification does affect the amplitude of the waves, which (for the case of the velocity perturbations) increases or decreases as they propagate from denser to lighter regions or viceversa \citep{Ferraro1958ApJ...127..459F,Zhugzhda1984A&A...132...45Z,Cargill1997ApJ...488..854C}. In the case of partially ionized plasmas, the numerical results of \citet{Popescu2019A&A...627A..25P,Popescu2019A&A...630A..79P} have shown that the damping caused by charge-neutral collisions competes with the amplitude growth due to the stratification. The general validity of neglecting the effect of gravity in the present study is briefly discussed in Appendix \ref{sec:strat}.

    It is also interesting to note that Eqs. (\ref{eq:lin_presn}) and (\ref{eq:lin_presc}) do not include either contribution from the energy transfer terms given by Eqs. (\ref{eq:qn}) and (\ref{eq:qc}). The second factor of $Q_{\rm{nc}}$ and $Q_{\rm{cn}}$ has a quadratic dependence on the drift velocity between the two fluids; therefore, it is straightforwardly disregarded by the linear approximation. However, the first factor in those expressions depends linearly on the difference of temperatures and it could be consistently retained by this approximation. However, it has also been neglected. The reason is that those factors tend to remove the differences in temperatures in a very short timescale given by
    \begin{equation} \label{eq:tau_temps}
        \tau_{\rm{coll}} = \frac{1}{\nu_{\rm{nc}} + \nu_{\rm{cn}}},
    \end{equation}
    as shown by \citet{Oliver2016ApJ...818..128O}. A similar result was obtained by \citet{Spitzer1956pfig.book.....S} for the case of collisions between ions and electrons in a fully ionized plasma. Therefore, the contribution of the energy transfer terms can be safely overlooked by the present linear analysis.

\subsection{Normal mode analysis} \label{sec:normal_modes}
    Here, we derive the dispersion relation that describes the properties of Alfvénic waves in a partially ionized plasma under the physical conditions detailed in the previous section. We perform a Fourier analysis in space and a normal mode analysis in time by assuming that the spatial and temporal dependence of the small amplitude perturbations is given by $\exp \left(i k_{z}z - i \omega t \right)$, where $k_{z}$ is the wavenumber along the direction of propagation and $\omega$ is the angular frequency of the wave. We take also into account that Alfvén waves are incompressible \citep[see, e.g.,][]{Cramer2001paw..book.....C,Goossens2003ASSL..294.....G,Goedbloed2004prma.book.....G}, so $\nabla \cdot \bm{V}_{\rm{n1}} = \nabla \cdot \bm{V}_{\rm{c1}} = 0$, and the continuity and pressure equations, that is Eqs. (\ref{eq:lin_rho_n}), (\ref{eq:lin_rho_c}), (\ref{eq:lin_presn}), and (\ref{eq:lin_presc}), can be dropped from the current computation. Furthermore, Alfvén waves are polarized in the direction transverse to the magnetic field (which we have assumed oriented along the $z$ direction), so we consider only the perturbations of velocity and magnetic field along the $x$ and $y$ directions.

    Then, from Eqs. (\ref{eq:lin_momn}), (\ref{eq:lin_momc}), and (\ref{eq:lin_induc}) we get to a system of 6 equations where the $x$ and $y$ components are coupled by the presence of Hall's term in the induction equation. Without the contribution of Hall's term, we would get an independent system of 3 equations for each component that would result in an equivalent dispersion relation to the one derived by \citet{Soler2013ApJ...767..171S}, who considered the vorticity perturbations of the plasma as the reference variables instead of the velocities. That dispersion relation describes the properties of linearly polarized Alfvén waves in a partially ionized plasma. However, when Hall's term is at play, Alfvénic perturbations are no longer linearly polarized but circularly polarized \citep{Lighthill1960RSPTA.252..397L,Stix1992wapl.book.....S,Cramer2001paw..book.....C}. Consequently, instead of employing the $x$ and $y$ components of the variables for the analysis, it is more useful to resort to the corresponding circularly polarized variables, defined as follows:
    \begin{equation} \label{eq:circ_vars}
        V_{\rm{s\pm}} = V_{\rm{s}x} \pm i V_{\rm{s}y}, \quad
        B_{\rm{1\pm}} = B_{\rm{1}x} \pm i B_{\rm{1}y},
    \end{equation}
    with $s \in \{c,n\}$. Here, the ``+'' sign corresponds to the left-hand polarization ($L$) and the ``-'' sign to the right-hand polarization ($R$).

    Taking into account the definitions given by Eq. (\ref{eq:circ_vars}), we arrive at the following set of equations:
    \begin{equation} \label{eq:vn_pm}
        \omega_{\pm} V_{\rm{n\pm}} = i \nu_{\rm{nc}} \left(V_{\rm{c\pm}} - V_{\rm{n\pm}} \right),
    \end{equation}

    \begin{equation} \label{eq:vc_pm}
        \omega_{\pm} V_{\rm{c\pm}} = -\frac{k_{z\pm} B_{\rm{0}}}{\mu_{\rm{0}} \rho_{\rm{c0}}} B_{\rm{1\pm}} + i \nu_{\rm{cn}} \left(V_{\rm{n\pm}} - V_{\rm{c\pm}} \right)
    \end{equation}

    \begin{equation} \label{eq:b1_pm}
        \omega_{\pm} B_{\rm{1\pm}} = - k_{z\pm} B_{\rm{0}} V_{\rm{c\pm}} \mp \frac{k_{z\pm}^{2}B_{\rm{0}}}{e n_{\rm{i0}} \mu_{\rm{0}}} B_{\rm{1\pm}},
    \end{equation}
    where it has been taken into account that in a hydrogen plasma $n_{\rm{e0}} = n_{\rm{i0}}$. The dispersion relation is finally obtained by combining the three previous equations and is given by
    \begin{gather} 
        \omega_{\pm}^{3} + i \left(\nu_{\rm{cn}} + \nu_{\rm{nc}} \right) \omega_{\pm}^{2} \pm \frac{k_{z\pm}^{2}c_{\rm{A}}^{2}}{\Omega_{\rm{i}}} \omega_{\pm}^{2} -k_{z\pm}^{2} c_{\rm{A}}^{2} \omega_{\pm}  \nonumber \\
        \pm i \left(\nu_{\rm{cn}}  + \nu_{\rm{nc}} \right) \frac{k_{z\pm}^{2} c_{\rm{A}}^{2}}{\Omega_{\rm{i}}}\omega_{\pm} - i k_{z\pm}^{2} c_{\rm{A}}^{2} \nu_{\rm{nc}} = 0,
        \label{eq:alfven_dr_hall}
    \end{gather}
    where $c_{\rm{A}}$ is the Alfvén speed defined in terms of the density of the charged fluid only as
    \begin{equation} \label{eq:cA}
        c_{\rm{A}} = \frac{B_{\rm{0}}}{\sqrt{\mu_{\rm{0}} \rho_{\rm{c0}}}}.
    \end{equation}

    Therefore, for each circular polarization we have an independent dispersion relation which is a polynomial of third order in $\omega$ and of second order in $k_{z}$. This means that the number of possible wave modes depends on whether the perturbation is generated by an impulsive or by a periodic driver, respectively. As it will be checked later, when $\omega_{\pm} \ll \Omega_{\rm{i}}$ the solutions of the two different systems converge to the solutions for linearly polarized Alfvén waves. However, as the frequency increases, the properties of the waves strongly depend on their polarization state. In the range of high frequencies (with $\omega > 0$), the $L$ modes are commonly known as ion cyclotron waves, while the $R$ modes are usually referred to as ion whistler waves \citep{Cramer2001paw..book.....C}.
    
    In the following sections we will consider some specific limits of Eq. (\ref{eq:alfven_dr_hall}), obtain various approximate analytical solutions and compare with the exact solutions and full numerical results.
    
\section{Results} \label{sec:results}
\subsection{Alfvén waves} \label{sec:alfvén}
    We start this section by exploring the range of frequencies where $\omega_{\pm} \ll \Omega_{\rm{i}}$. In this limit, the dispersion relation given by Eq. (\ref{eq:alfven_dr_hall}) simplifies to
    \begin{equation} \label{eq:alfven_dr}
        \omega^{3} + i \left(1+ \chi \right) \nu_{\rm{nc}} \omega^{2} - k_{z}^{2} c_{\rm{A}}^{2} \omega - i\nu_{\rm{nc}} k_{z}^{2} c_{\rm{A}}^{2} = 0,
    \end{equation}
    where the relation $\chi = \nu_{\rm{cn}} / \nu_{\rm{nc}}$ has been taken into account and the subscripts ``$\pm$'' have been dropped since there is no distinction between the two circular polarizations. This is the same formula that has already been derived by \citet{Soler2013ApJ...767..171S}, which also agrees with the respective expressions from the works of \citet{Kulsrud1969ApJ...156..445K}, \citet{DePontieu1998A&A...338..729D}, \citet{Kumar2003SoPh..214..241K}, \citet{Zaqarashvili2011A&A...529A..82Z} or \citet{Mouschovias2011MNRAS.415.1751M}, among others.

    Since we are interested here in the study of the propagation of waves generated by a periodic driver, it is convenient to rewrite the dispersion relation as
    \begin{equation} \label{eq:alfven_dr2}
        k_{z}^{2} = \frac{\omega^{2}}{c_{\rm{A}}^{2}}\frac{\omega + i \left(1 + \chi \right) \nu_{\rm{nc}}}{\omega + i \nu_{\rm{nc}}}.
    \end{equation}

    Then, we assume that the frequency $\omega$ is real while allowing the wavenumber to be complex, $k_{z} = k_{\rm{R}} + i k_{\rm{I}}$ (for the sake of simplicity, we have dropped the subscript ``$z$'' for the real and imaginary parts of the wavenumber). A value of $k_{\rm{I}}$ different from $0$ means that the waves are damped in space. If $\omega > 0$, positive (negative) values of $k_{R}$ represent forward (backward) propagating waves, that is waves propagating in the positive (negative) direction along the $z$ axis. As shown by \citet{Soler2013ApJ...767..171S}, exact analytical solutions can be obtained for the real and imaginary parts of $k_{z}$. To do so, we insert the expression $k_{z} = k_{\rm{R}} + i k_{\rm{I}}$ into Eq. (\ref{eq:alfven_dr2}) and solve the resulting system of two equations for $k_{\rm{R}}^{2}$ and $k_{\rm{I}}^{2}$. Thus, we obtain the following solutions:
    \begin{eqnarray} \label{eq:kzr2_exact}
        k_{\rm{R}}^{2} &=& \frac{1}{2}\frac{\omega^{2}}{c_{\rm{A}}^{2}} \frac{\omega^{2}+ \left(1 + \chi \right) \nu_{\rm{nc}}^{2}}{\omega^{2}+\nu_{\rm{nc}}^{2}} \nonumber \\
        &\times& \left[1 + \left(1 + \frac{\chi^{2} \nu_{\rm{nc}}^{2} \omega^{2}}{\left(\omega^{2} + \left(1 + \chi \right) \nu_{\rm{nc}}^{2} \right)^{2}} \right)^{1/2} \right],
    \end{eqnarray}
    and
    \begin{equation} \label{eq:kzI2_exact}
        k_{\rm{I}}^{2} = k_{\rm{R}}^{2} - \frac{\omega^{2}}{c_{\rm{A}}^{2}} \frac{\omega^{2} + \left(1 + \chi \right) \nu_{\rm{nc}}^{2}}{\omega^{2} + \nu_{\rm{nc}}^{2}}.
    \end{equation}

    Although the expressions above provide the exact solutions of the dispersion relation, it is also useful to find simpler analytical approximations that can be applied to certain limits. In the next section we compute and check the validity of several expressions for the limits of weak and strong collisional coupling, that is $\nu_{\rm{nc}} \ll \omega$ and $\nu_{\rm{nc}} \gg \omega$, respectively.
    
\subsubsection{Comparison of approximate and exact solutions}
    Following the procedure by \citet{Soler2013ApJ...767..171S}, to find the approximate solutions in the limits of weak or strong coupling we perform Taylor series expansions to Eqs. (\ref{eq:kzr2_exact}) and (\ref{eq:kzI2_exact}) assuming that $\nu_{\rm{nc}} \to 0$ or $\omega \to 0$, respectively, and retain the first non-zero terms only.

    For the case of weak coupling (which will be denoted by the subscript ``W''), when $\nu_{\rm{nc}} \ll \omega$, the real part of the wavenumber is given by
    \begin{equation} \label{eq:kr_weak}
        k_{\rm{R,W}} \approx \pm \frac{\omega}{c_{\rm{A}}},
    \end{equation}
    recovering the expression for a fully ionized plasma, and the imaginary part is
    \begin{equation} \label{eq:ki_weak}
        k_{\rm{I,W}} \approx \pm \frac{\chi \nu_{\rm{nc}}}{2 c_{\rm{A}}} = \pm \frac{\nu_{\rm{cn}}}{2 c_{\rm{A}}},
    \end{equation}
    which shows that in this limit the damping rate does not depend on the frequency of the wave.

    \begin{figure*}
        \centering
        \includegraphics[width=0.49\hsize]{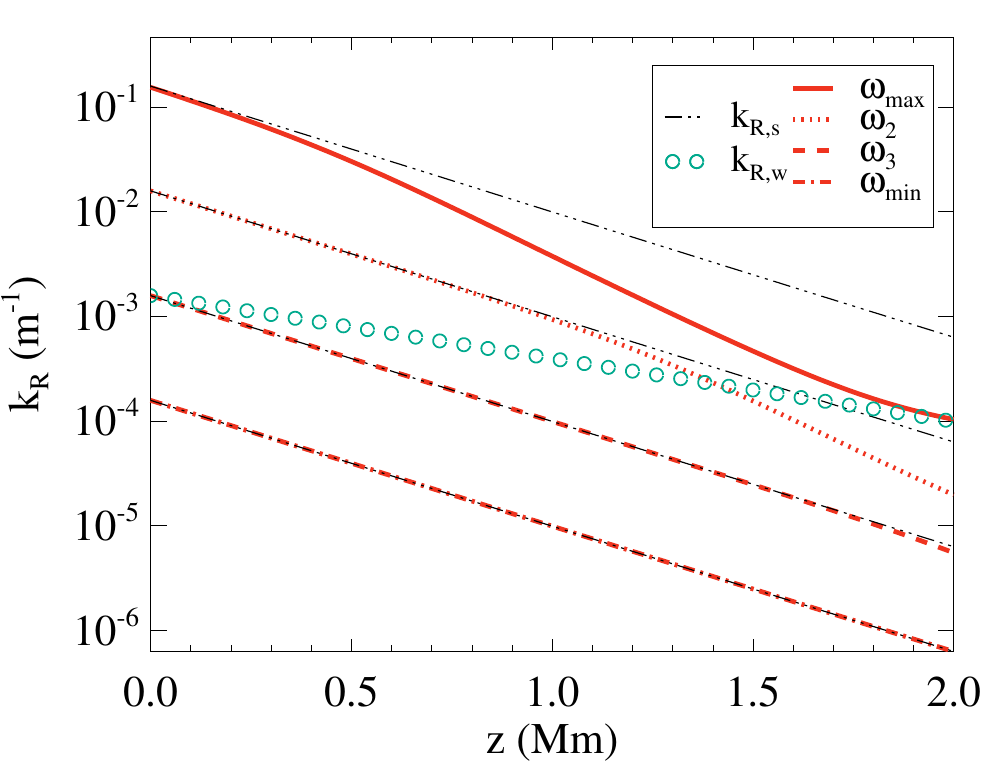}
        \includegraphics[width=0.49\hsize]{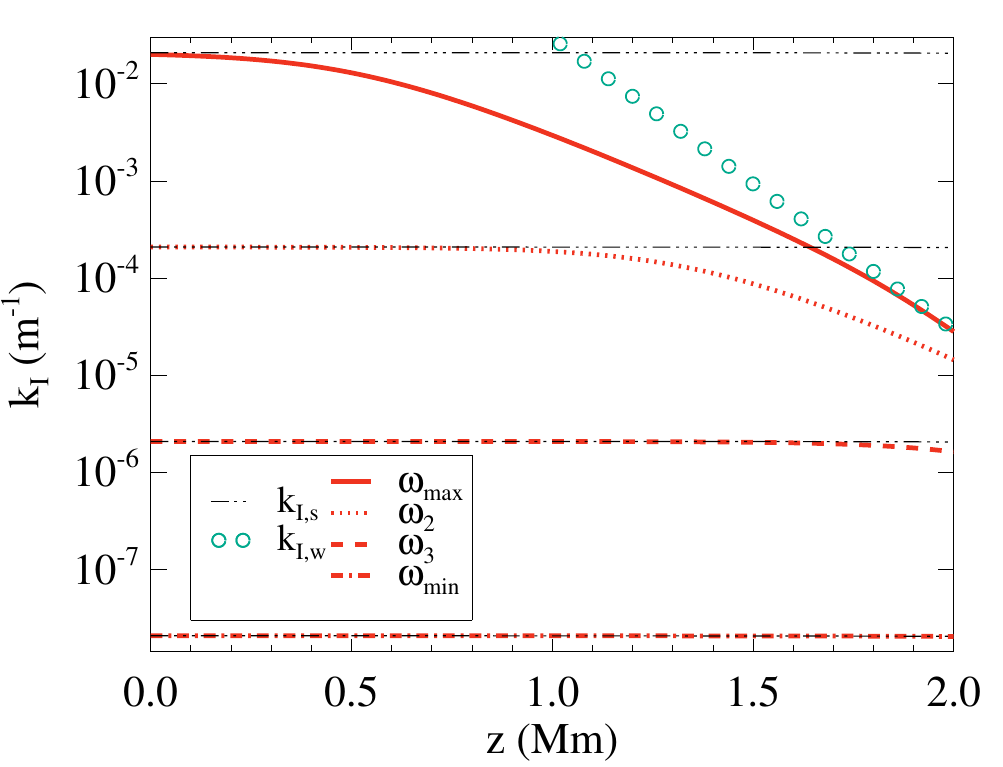}
        \caption{Real part (left panel) and imaginary part (right panel) of the wavenumber of Alfvén waves as functions of height. Exact solutions of the dispersion relation are represented by the red lines, with each linestyle corresponding to a different wave frequency from the set $\omega = \{0.02 \pi, 0.2 \pi, 2 \pi, 20\pi \} \ \rm{rad \ s^{-1}}$. The black dotted-dashed lines and the blue circles represent the approximate results from Eqs. (\ref{eq:kr_strong})-(\ref{eq:ki_strong}) and (\ref{eq:kr_weak})-(\ref{eq:ki_weak}), respectively.}
        \label{fig:kRkI_alfven}
    \end{figure*}

    On the other hand, for the case of strong coupling with $\nu_{\rm{nc}} \gg \omega$ (denoted by the subscript ``S''), the real part of the wavenumber is
    \begin{equation} \label{eq:kr_strong}
        k_{\rm{R,S}} \approx \pm \frac{\omega}{c_{\rm{A}}}\sqrt{1 + \chi} = \pm \frac{\omega}{c_{\rm{A,mod}}},
    \end{equation}
    where $c_{\rm{A,mod}}$ is the modified Alfvén speed defined in terms of the total density of the plasma, that is taking also into account the contribution from the neutral fluid \citep[see, e.g.,][]{Soler2013ApJ...767..171S,Zaqarashvili2013A&A...549A.113Z}, and the imaginary part is
    \begin{equation} \label{eq:ki_strong}
        k_{\rm{I,S}} \approx \pm \frac{\chi \omega^{2}}{2 \sqrt{1+\chi}c_{\rm{A}}\nu_{\rm{nc}}},
    \end{equation}
    which shows that the damping rate becomes inversely proportional to the collision frequency $\nu_{\rm{nc}}$ but proportional to the square of the wave frequency\footnote{We note here that the published version of \citet{Soler2013ApJ...767..171S} contained some typos in their Eq. (42) regarding the value of $k_{\rm{I,S}}$, and we show in our Eq. (\ref{eq:ki_strong}) the corrected formula.}

    After showing the expressions for the analytical approximations, we now turn our attention to the comparison with the exact solutions of Eq. (\ref{eq:alfven_dr}) under the physical conditions of the atmospheric model described in Section {\ref{sec:background}}. This comparison is illustrated in Fig. \ref{fig:kRkI_alfven}, where the real part and the imaginary part of the results are shown on the left and the right panels, respectively. Due to the symmetry of the solutions from the dispersion relation, only those with $k_{\rm{R}} \geq 0$ are displayed. The red lines correspond to the exact solutions given by Eqs. (\ref{eq:kzr2_exact}) and (\ref{eq:kzI2_exact}), with each different linestyle associated to a particular value of the wave frequencies considered in this study (see Section \ref{sec:background}). The approximations for the strong coupling limit, provided by Eqs. (\ref{eq:kr_strong}) and (\ref{eq:ki_strong}), are represented by the black dotted-dashed lines, while the approximations for the weak coupling limit, Eqs. (\ref{eq:kr_weak}) and (\ref{eq:ki_weak}), are shown as circles. For the sake of clarity and to avoid the presence of too many lines in the plots, the latter approximations have only been included for the case $\omega_{\rm{max}}$, since it is the only scenario where they are in good agreement with the exact solutions in at least a small region of the atmosphere.

    The left panel of Fig. \ref{fig:kRkI_alfven} shows that for any wave frequency $\omega$, the wavenumber of the perturbations decreases as the height increases (and, consequently, the wavelength becomes larger, as described in Appendix \ref{sec:wavelengths}). This is caused by the growth of the Alfvén speed as the plasma becomes less dense. At the bottom layers of the atmosphere the strong coupling approximations and the exact solutions are in very good agreement for all the considered frequencies and $k_{\rm{R}}$ is directly proportional to $\omega$. However, at the upper layers the strong coupling limit remains accurate only for the waves with larger periods (or lower frequencies, $\omega_{\rm{min}}$ and $\omega_{\rm{3}}$). For the case of $\omega_{\rm{2}} = 2\pi \ \rm{rad \ s^{-1}}$, the actual wavenumber starts to deviate from the approximate value at the height $z \sim 1.2 \ \rm{Mm}$, where $\omega_{\rm{3}} \gtrsim \nu_{\rm{nc}}$. For the largest wave frequency, $\omega_{\rm{max}}$, the clear separation from the strong coupling limit begins at a height $z \sim 0.4 \ \rm{Mm}$. According to Fig. \ref{fig:freqs}, $\omega_{\rm{max}}$ becomes much larger than $\nu_{\rm{nc}}$ as height increases and we can see in Fig. \ref{fig:kRkI_alfven} that the exact solution of the dispersion relation tends to the weak coupling approximation.

    In the right panel of Fig. \ref{fig:kRkI_alfven} we can check the quadratic dependence on $\omega$ of the approximations for the strong coupling range given in Eq. (\ref{eq:ki_strong}), so the damping due to charge-neutral collisions is much larger for high-frequency waves than for low-frequency waves. In addition, we notice the surprising behavior that these approximate results do not vary with height. This is a consequence of the particular set of parameters chosen to define the background atmosphere (where the magnetic field and the temperature are uniform). Then, we see again that at the middle and upper layers of the atmosphere the strong coupling approximations become inaccurate for the higher frequency waves. For this range of frequencies the damping rates decrease with height, varying by several orders of magnitude. Taking into account Eq. (\ref{eq:ki_weak}), this steep decrease comes from the combination of the large reduction of the charge-neutral collision frequency, $\nu_{\rm{cn}}$, represented in Fig. \ref{fig:freqs}, and the increase of the Alfvén speed.

    In Fig. \ref{fig:kRkI_alfven} we have presented the properties of Alfvén waves in terms of the real and imaginary parts of the wavenumber because these parameters are obtained directly from the dispersion relation. However, in some situations it may be more informative to describe the waves in terms of their wavelength and damping lengths, as it could happen, for instance, when comparisons with observations are performed. Since both approaches yield equivalent conclusions, it is not worth to include those additional results in the current section, but we provide them in Appendix \ref{sec:wavelengths}.

\subsubsection{Quality factor and phase speed}
    Once we have the results from the dispersion relation, we can compute two other parameters that provide useful information about the properties of the waves: the quality factor and the phase speed. The former gives a measure of the relative strength of the damping of the waves in comparison with their wavenumbers or their frequencies, while the latter represents the speed at which the waves propagate.
    
    For the case of perturbations generated by a periodic driver, a common definition of the quality factor is
    \begin{equation} \label{eq:q_k}
        Q_{(k)} = \frac{1}{2}\frac{|k_{\rm{R}}|}{|k_{\rm{I}}|}.
    \end{equation}
    For values $Q_{(k)} > 1/2$, the waves are said to be underdamped and they show an oscillatory behavior with an amplitude that decreases as the perturbation propagates; in the limit $Q_{(k)} \to \infty$ there is no damping ($k_{\rm{I}} = 0$). For $Q_{(k)} < 1/2$ the waves are overdamped, meaning that there is a very strong damping and the propagation of the perturbations is limited to very short distances; in the limit $Q_{(k)} = 0$, corresponding to $k_{\rm{R}} = 0$, the perturbations are evanescent and there is no propagation.

    \begin{figure}
        \includegraphics[width=\hsize]{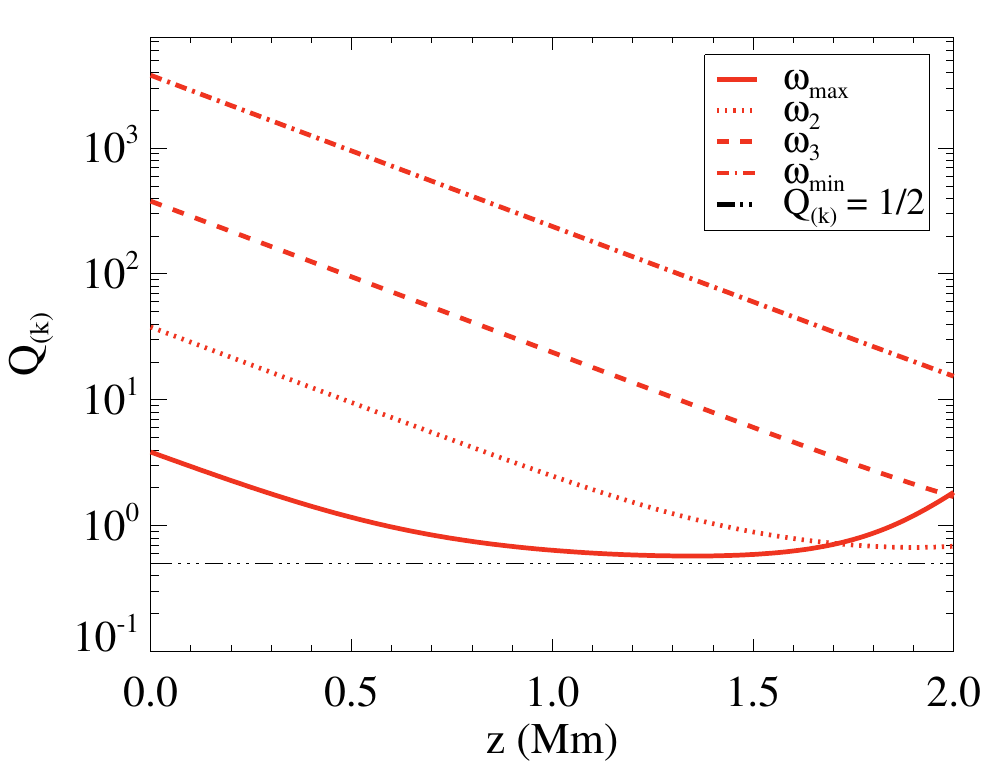}
        \caption{Quality factor of Alfvén waves as a function of height. The red lines represent the results for the different values of wave frequency, following the style from Fig. \ref{fig:kRkI_alfven}. The horizontal black line represents the critical value $Q_{(k)} = 1/2$.}
        \label{fig:qk_alfven}
    \end{figure}

    We present in Fig. \ref{fig:qk_alfven} the dependence on height of the quality factor for the solutions of the dispersion relation already analyzed in Fig. \ref{fig:kRkI_alfven}. We see that the lower frequency waves ($\omega_{\rm{min}}$ and $\omega_{\rm{3}}$) are clearly underdamped in the whole atmosphere, with $Q_{(k)} \gg 1/2$, and the quality factor decreases with height. Comparing with Fig. \ref{fig:kRkI_alfven}, we see that this fact is not caused by an increase of the damping rate ($k_{\rm{I}}$) but by the decrease of the wavenumber of the perturbations. In the case of $\omega_{\rm{2}}$, we see that the resulting wave is always more efficiently damped than the low-frequency ones in the whole atmosphere, getting close to the critial value that separates the underdamped and overdamped regimes ($Q_{(k)} = 1/2$) at the upper layers. The more distinct behavior is again found for the largest frequency wave: its quality factor does not monotonically decrease with height in this atmosphere but reaches a minimum at the middle layers and then rises again at the top layers. Thus, we find that, in terms of the quality factor, Alfvén waves are more efficiently damped by charge-neutral collisions in the regime $\omega \sim \nu_{\rm{nc}}$, in good agreement with \citet{Leake2005A&A...442.1091L}, \citet{Zaqarashvili2011A&A...529A..82Z} and \citet{Soler2013ApJ...767..171S}. We note also that in none of the four cases studied here the quality factor reaches the region of overdamped waves ($Q_{(k)} < 1/2$), a similar result to that found by \citet{Soler2015A&A...573A..79S} using a different atmospheric model for the solar chromosphere.
    
    Then, according to the usual definitions of phase velocity and phase speed \citep[see, e.g.,][]{Cramer2001paw..book.....C,Goossens2003ASSL..294.....G,Goedbloed2004prma.book.....G}, and following \citet{Soler2013ApJ...767..171S}, the phase speed can be computed from Eq. (\ref{eq:alfven_dr2}) as
    \begin{equation} \label{eq:vphase}
        V_{\rm{ph}} = \frac{\omega}{k_{z}} = c_{\rm{A}} \sqrt{\frac{\omega + i \nu_{\rm{nc}}}{\omega + i \left(1 + \chi \right) \nu_{\rm{nc}}}}.
    \end{equation}
    Equivalently, it can be written as
    \begin{equation} \label{eq:vphase2}
        V_{\rm{ph}} = \frac{\omega}{k_{\rm{R}} + i k_{\rm{I}}} = \frac{\omega k_{\rm{R}}}{k_{\rm{R}}^{2} + k_{\rm{I}}^2} - i \frac{\omega k_{\rm{I}}}{k_{\rm{R}}^{2} + k_{\rm{I}}^{2}},
    \end{equation}
    where $k_{\rm{R}}$ and $k_{\rm{I}}$ are given by Eqs. (\ref{eq:kzr2_exact}) and (\ref{eq:kzI2_exact}). The real part of the phase speed is related to the actual propagation speed of the perturbations, while the imaginary part is related to the damping caused by the collisional interaction. It can be checked that in the weak and the strong coupling regimes, the phase speed is given by
    \begin{equation} \label{eq:vphase_lims}
        V_{\rm{ph}} \approx c_{\rm{A}} \quad \text{and} \quad V_{\rm{ph}} \approx \frac{c_{\rm{A}}}{\sqrt{1 + \chi}} = c_{\rm{A,mod}}, 
    \end{equation}
    respectively.

    \begin{figure}
        \includegraphics[width=\hsize]{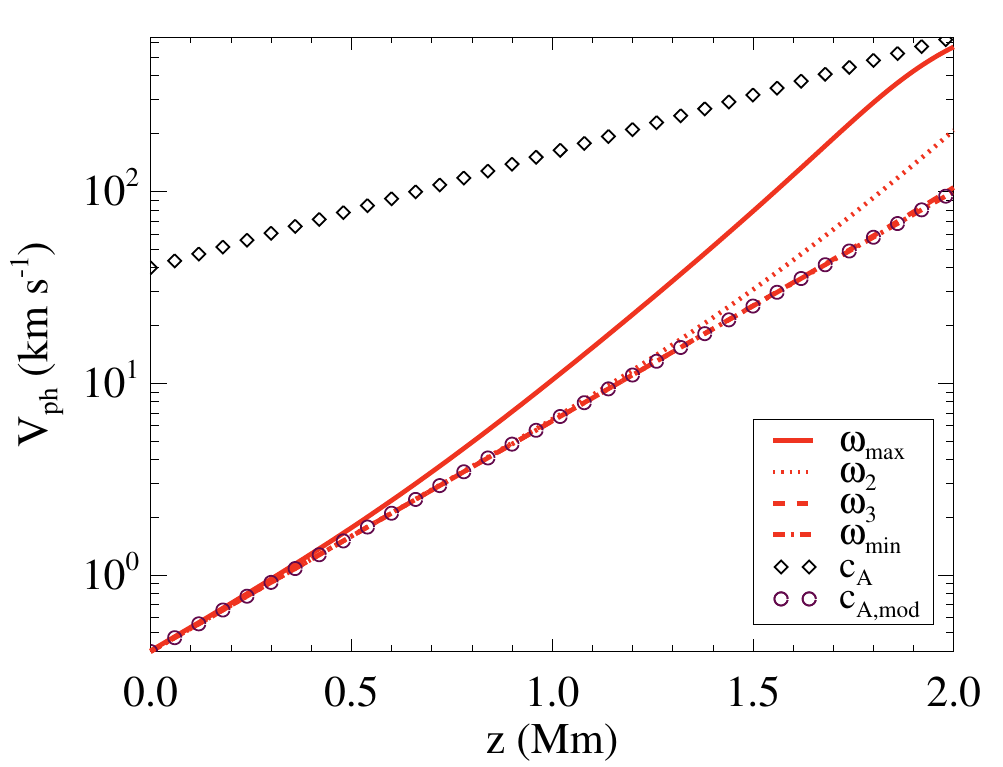}
        \caption{Real part of the phase speed of Alfvén waves as function of height. Diamonds and circles represent the classical and modified version of the Alfvén speed, $c_{\rm{A}}$ and $c_{\rm{A,mod}}$, respectively.}
        \label{fig:vph_alfven}
    \end{figure}

    Figure \ref{fig:vph_alfven} shows how the phase speeds of the considered Alfvén waves depend on height. For reference, we have also included in the plot the values of the classical Alfvén speed, $c_{\rm{A}}$, and its modified version that takes into account the density of neutrals, $c_{\rm{A,mod}}$: both speeds increase with height due to the drop in density of the charged fluid and of the whole plasma. Due to the strong collisional coupling between the two fluids, the low-frequency perturbations propagate at the modified Alfvén speed in the whole atmosphere. On the other hand, the propagation speed of the high-frequency waves strongly varies with height and finally tends to the Alfvén speed of the fully ionized scenario at the top layers.

\subsubsection{Eigenfunctions relations} \label{sec:eigen_alfvén}
     Up to this point of the investigation we have only focused on the properties of the normal modes that can be directly extracted from the solutions of Eq. (\ref{eq:alfven_dr}). Nonetheless, it is also informative to study the relations between the eigenfunctions associated to those normal modes, also refereed to as polarization relations \citep[see, e.g.,][]{Goedbloed2004prma.book.....G,Walker2004SPP....16.....W,Khomenko2012ApJ...746...68K,Popescu2019A&A...627A..25P}. For the case of linearly polarized Alfvén waves, the relevant eigenfunction relations can be obtained from Eqs. (\ref{eq:vn_pm})-(\ref{eq:b1_pm}) by neglecting the contribution of the Hall term in the induction equation.

     In the first place, we pay attention to the relation between the velocity perturbations. Therefore, from Eq. (\ref{eq:vn_pm}) we find that
     \begin{equation} \label{eq:r_vcvn}
        R_{\rm{V}} \equiv \frac{V_{\rm{c}}}{V_{\rm{n}}} = \frac{-i \omega + \nu_{\rm{nc}}}{\nu_{\rm{nc}}}.
     \end{equation}
    An equivalent but more convoluted expression can be derived from the combination of the two remaining equations. This expression is also equivalent to Eq. (38) of \citet{Popescu2019A&A...627A..25P} after applying the appropriate change of notation and performing the transformation $\omega \to -\omega$, since those authors assumed a dependence on $\exp \left(i \omega t\right)$ instead of on $\exp \left(-i \omega t\right)$ in their normal mode analysis.

    \begin{figure*}
        \centering
        \includegraphics[width=0.49\hsize]{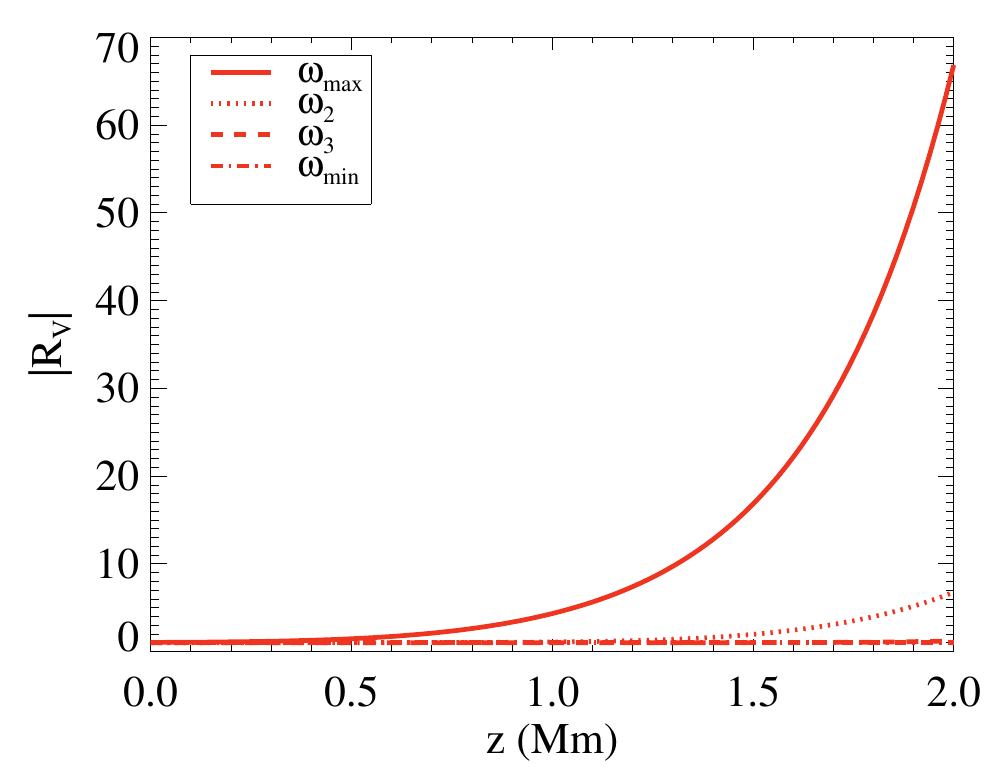}
        \includegraphics[width=0.49\hsize]{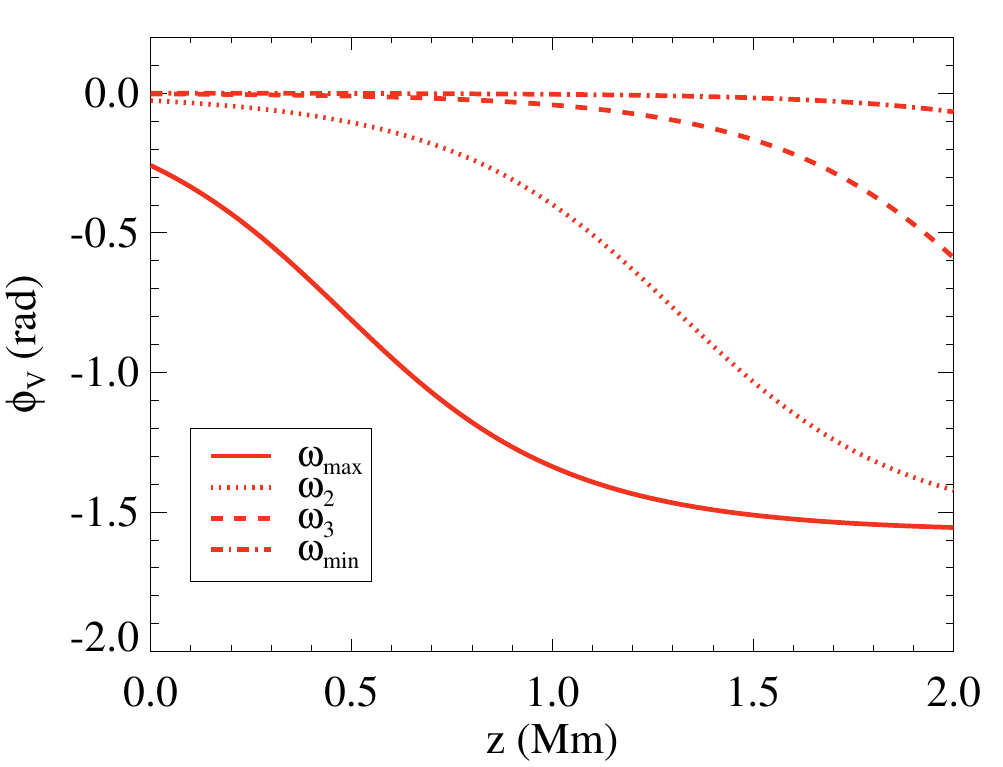}
        \caption{Ratio of charges to neutrals velocities for Alfvén waves in a weakly ionized atmosphere. The modulus of the ratio, $|R_{\rm{V}}|$, is displayed in the left panel, while the phase shift, $\Phi_{\rm{V}}$, is shown on the right panel.}
        \label{fig:ratio_vels}
    \end{figure*}

    In the strong coupling limit ($\nu_{\rm{nc}} \gg \omega $) we find the expected result that $R_{\rm{V}} \approx 1$ and the amplitude of the velocities of the two components of the plasma is the same. On the opposite limit ($\nu_{\rm{cn}} \ll \omega$), the ratio $R_{\rm{V}}$ grows without bound, which means that the neutral fluid is hardly affected by the perturbation. This result is consistent with the multi-fluid numerical simulations performed by \citet{MartinezGomez2017ApJ...837...80M}, who show in their Figure 14 how the relative amplitudes of the velocities of neutrals decrease as the frequency of the Alfvén wave increases. In the intermediate range of coupling, $R_{\rm{V}}$ is a complex quantity with both its real and imaginary parts different from zero. Therefore, a phase shift between the charges and the neutrals velocities appears. To study this behavior it is convenient to rewrite the velocity ratio as
    \begin{equation} \label{eq:r_vcvn2}
        R_{\rm{V}} = |R_{\rm{V}}| \exp \left(i \Phi_{\rm{V}} \right),
    \end{equation}
    where $\Phi_{\rm{V}}$ represents the phase difference.

    The modulus of the ratio of velocities and the phase shift as functions of height are shown in Fig. \ref{fig:ratio_vels} for the chosen reference frequencies. We see in the left panel that $|R_{\rm{v}}| \approx 1$ for the low-frequency waves in the whole atmosphere, but only at the bottom region for the high-frequency waves, whose ratios increase with height. On the right panel of Fig. \ref{fig:ratio_vels}, the phase shift is $\Phi_{\rm{V}} \approx 0$ for all the frequencies (except for the largest one) at the bottom region. Then, the absolute value of the phase shift increases with height. The largest deviation from $\Phi_{\rm{V}} = 0$ is found for $\omega_{\rm{max}}$, with a phase shift that tends to the value $\Phi_{\rm{V}} \approx -\pi/2$ at the top of the atmosphere. The existence of these phase shifts in propagating Alfvén waves when the collisional coupling is not strong enough has already been reported in the numerical studies of \citet{MartinezGomez2017ApJ...837...80M,MartinezGomez2018ApJ...856...16M}, \citet{Popescu2019A&A...627A..25P} and \citet{MartinezSykora2020ApJ...900..101M}. Similar results have been found for the different components of multi-ion plasmas in the appropriate regime of wave frequencies \citep{MartinezGomez2016ApJ...832..101M,MartinezSykora2020ApJ...900..101M}. 
    
    We remark here that the variation of the relative amplitudes of the velocities of charges and neutrals and the presence of phase shifts are not exclusive features of propagating waves. They have also been found in numerical simulations of standing Alfvén waves \citep[see, e.g.,][]{Soler2013ApJ...767..171S} 

    Now, from Eq. (\ref{eq:b1_pm}) we can obtain the following relation between the perturbations of magnetic field and the velocity of charges:
    \begin{equation} \label{eq:b1_vc}
        B_{\rm{1}} = - \frac{k_{z} B_{\rm{0}}}{\omega}V_{\rm{c}} = \mp \frac{\left(|k_{\rm{R}}| + i |k_{\rm{I}}|\right) B_{\rm{0}}}{\omega} V_{\rm{c}}, 
    \end{equation}
    where the two possible solutions (positive and negative) from the dispersion relation have been taken into account. Here, the sign ``-'' corresponds to waves propagating parallel to the background magnetic field, while the sign ``+'' corresponds to those propagating in the anti-parallel direction. In the case with no charge-neutral collisions, so $k_{\rm{I}} = 0$, we recover the classical relation
    \begin{equation} \label{eq:b1_vc_2}
        B_{\rm{1}} = \mp \frac{B_{\rm{0}}}{c_{\rm{A}}}V_{\rm{c}},
    \end{equation}
    which shows that, for propagating Alfvén waves, the perturbations of magnetic field and velocity of the ionized fluid are always completely in phase or in anti-phase \citep{Walen1944ArMAF..30A...1W,Priest1984smh..book.....P}. In the strong coupling limit, we find the same relation but substituting the classical Alfvén speed, $c_{\rm{A}}$, by its modified version, $c_{\rm{A,mod}}$.

    Since the ratio $B_{\rm{1}} / V_{\rm{c}}$ would compare quantities with different physical units, it is more convenient for the present study to define a non-dimensional ratio as follows:
    \begin{equation} \label{eq:r_b1vc}
        R_{\rm{B,Vc}} \equiv \frac{B_{\rm{1}}}{B_{\rm{0}}} \frac{c_{\rm{A,mod}}}{V_{\rm{c}}} \Rightarrow R_{\rm{B,Vc}} = \mp \frac{|k_{\rm{R}}| + i |k_{\rm{I}}|}{\omega} c_{\rm{A,mod}},
    \end{equation}
    which produces the result $R_{\rm{B,Vc}} \approx \mp 1$ in the strong coupling regime, and $R_{\rm{B,Vc}} \approx \mp 1 / \sqrt{1 + \chi}$ in the weak coupling limit. We note that the application of the normalization factor does not affect the phase shift between the two variables of interest, which will be given by $\Phi_{\rm{B,Vc}}$ after expressing $R_{\rm{B,Vc}}$ in a similar way to Eq. (\ref{eq:r_vcvn2}).

    \begin{figure*}
        \centering
        \includegraphics[width=0.49\hsize]{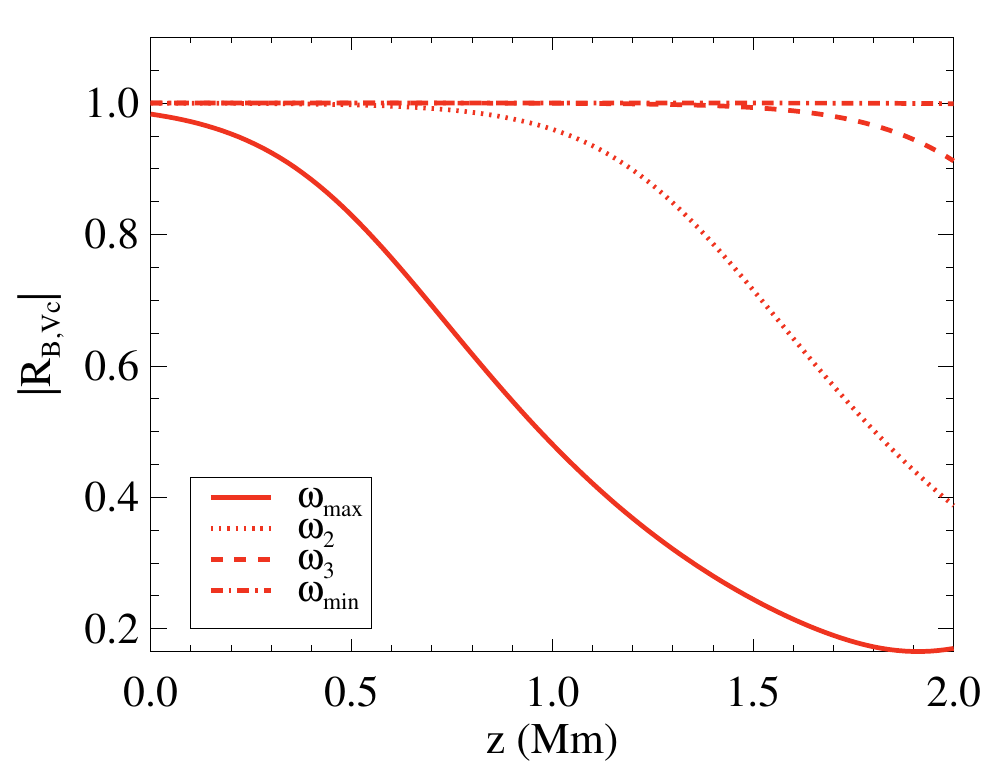}
        \includegraphics[width=0.49\hsize]{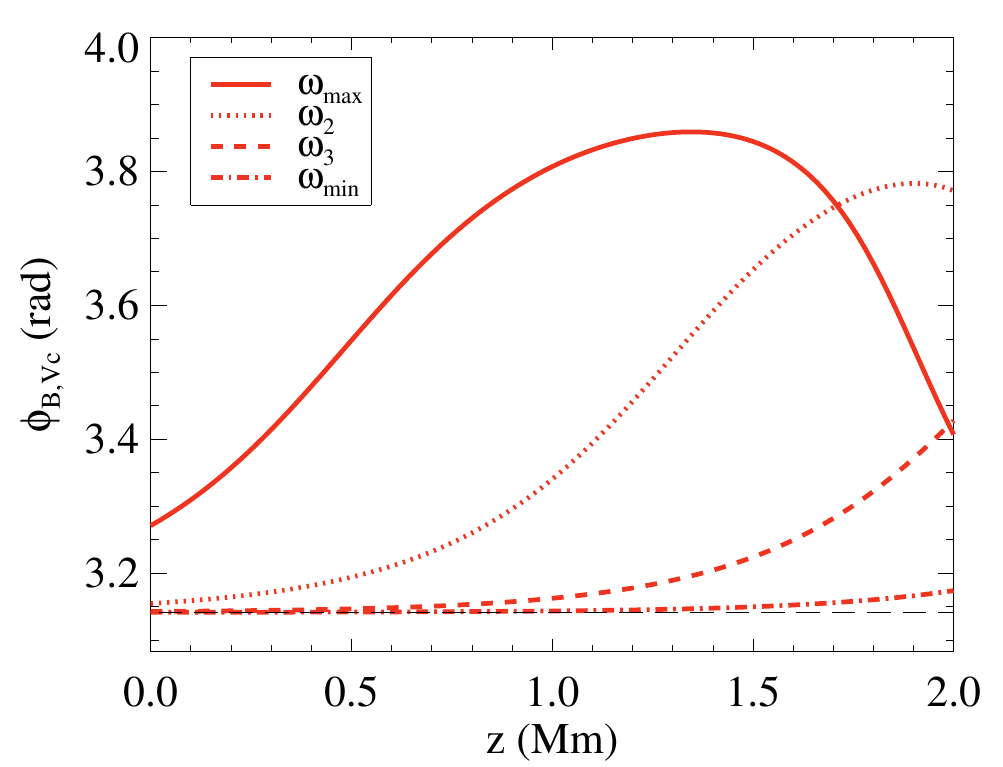}
        \caption{Non-dimensional ratio between perturbations of magnetic field and velocity of the charged fluid for the case of forward propagating Alfvén waves. The modulus of the ratio, $|R_{\rm{B,Vc}}|$, and the phase shift, $\Phi_{\rm{B,Vc}}$, are given on the left and right panels, respectively. The horizontal black dashed line in the right panel represents the phase angle $\Phi_{\rm{B,Vc}} = \pi$.}
        \label{fig:ratio_bvc}
    \end{figure*}

    \begin{figure*}
        \centering
        \includegraphics[width=0.49\hsize]{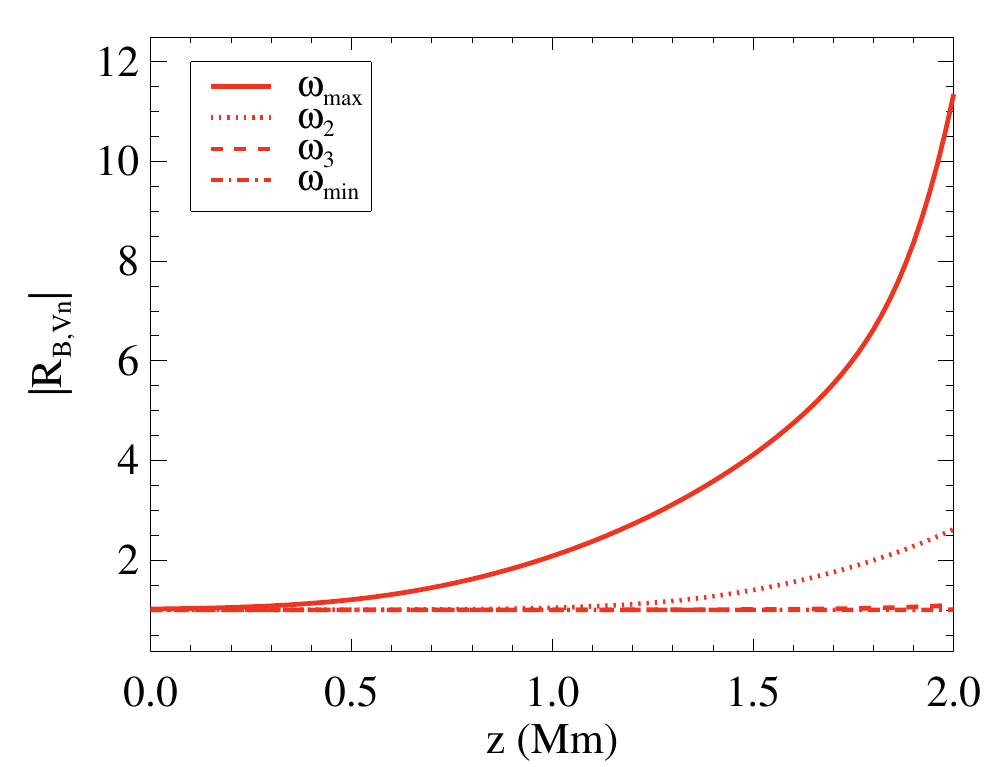}
        \includegraphics[width=0.49\hsize]{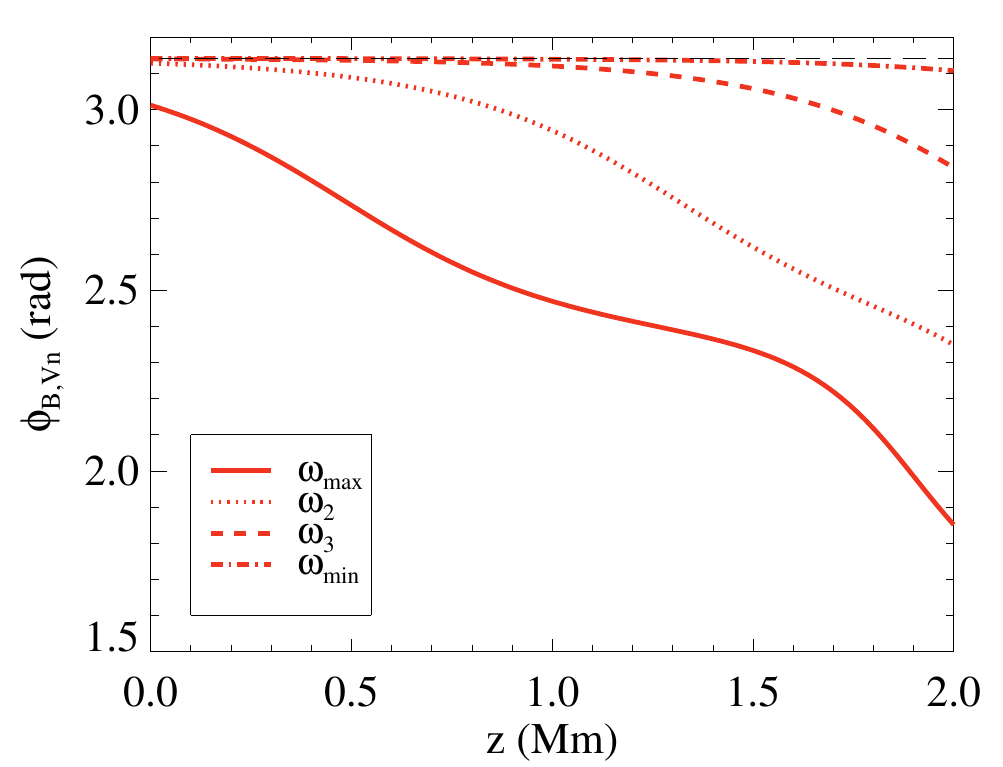}
        \caption{Non-dimensional ratio between perturbations of magnetic field and velocity of the neutral fluid for the case of forward propagating Alfvén waves. The modulus of the ratio, $|R_{\rm{B,Vn}}|$, and the phase shift, $\Phi_{\rm{B,Vn}}$, are given on the left and right panels, respectively. The horizontal black dashed line in the right panel represents the phase angle $\Phi_{\rm{B,Vn}} = \pi$.}
        \label{fig:ratio_bvn}
    \end{figure*}
    
    We present in Fig. \ref{fig:ratio_bvc} the dependence on height of the ratio $R_{\rm{B,Vc}}$ for the case of Alfvén waves propagating parallel to the background magnetic field. We see in the left panel the expected behavior (described in the previous paragraph) for the modulus of the ratio: on the one hand, $|R_{\rm{B,Vc}}| \approx 1$ for the wave with the lowest frequency, and in the regions where $\nu_{\rm{nc}} \gg \omega$ (see, Fig. \ref{fig:freqs}); on the other hand, $|R_{\rm{B,Vc}}| \approx 0.15$ for the shortest period wave ($\omega_{\rm{max}}$) at the top of the atmosphere where the collisional coupling is weak, in good agreement with the value of $1/\sqrt{1 + \chi}$ at that height.

    More surprising is the behavior of the phase shift $\Phi_{\rm{B,Vc}}$ depicted in the right panel of Fig. \ref{fig:ratio_bvc}. Here, it is clear that the phase shift departs from the value $\Phi_{\rm{B,Vc}} = \pi$ as the frequency of the waves moves away from the limits of strong and weak coupling. This means that in scenarios with an intermediate degree of collisional coupling, the perturbations of magnetic field and velocity of charges are no longer completely in phase opposition. Under the physical conditions studied in this work, the maximum phase shift reached by the high-frequency waves is $\Phi_{\rm{B,Vc}} \approx 5 \pi /4$. To the best of our knowledge, this variation of the phase shift between the magnetic field and the velocity has not been previously investigated. In Section \ref{sec:concl} we discuss its possible implications for the observation and identification of Alfvén waves in partially ionized plasmas.

    Although not shown here, we have checked that equivalent results regarding $R_{\rm{B,Vc}}$ are obtained for Alfvén waves propagating in the opposite direction of the background magnetic field. In that case, the phase shift between $B_{\rm{1}}$ and $V_{\rm{c}}$ varies from $\Phi_{\rm{B,Vc}} = 0$ to $\Phi_{\rm{B,Vc}} \approx \pi/4$.

    Finally, for the sake of completeness, we also analyze the relation between the perturbations of the magnetic field and the velocity of the neutral fluid. A similar ratio to that given by Eq. (\ref{eq:r_b1vc}) can be obtained through the combination of Eqs. (\ref{eq:vn_pm}) and (\ref{eq:b1_pm}). Thus, the following expression can be obtained:
    \begin{equation} \label{eq:r_b1vn}
        R_{\rm{B,Vn}} = \mp \frac{|k_{\rm{R}}| + i |k_{\rm{I}}|}{\omega} \left(\frac{-i \omega + \nu_{\rm{nc}}}{\nu_{\rm{nc}}} \right) c_{\rm{A,mod}},
    \end{equation}
    which fulfills that $R_{\rm{B,Vn}} = R_{\rm{B,Vc}} R_{\rm{V}}$. With this definition, in the strong coupling limit we have that $R_{\rm{B,Vn}} = \mp 1$, the same result as for $R_{\rm{B,Vc}}$. In contrast, in the weak coupling regime, we get that $R_{\rm{B,Vn}} \to \infty$, a behavior that is inherited from the ratio $R_{\rm{V}}$.

    The variation with height of the parameter $R_{\rm{B,Vn}}$ in the model atmosphere is represented in Fig. \ref{fig:ratio_bvn}. In the left panel, we see that the modulus of this ratio follows a similar trend to the one depicted for $|R_{\rm{V}}|$ in Fig. \ref{fig:ratio_vels}, although it reaches a lower maximum amplitude due to the contribution from $R_{\rm{B,Vc}}$. We can check that the relation $|R_{\rm{B,Vn}}| = |R_{\rm{B,Vc}}| |R_{\rm{V}}|$ is fulfilled.

    The right panel of Fig. \ref{fig:ratio_bvn} shows that $B_{\rm{1}}$ and $V_{\rm{n}}$ are in anti-phase ($\Phi_{\rm{B,Vn}} = \pi$) in the strong coupling regime, as expected. Then, in contrast with $\Phi_{\rm{B,Vc}}$ (which reaches a maximum value at the middle layers of the atmosphere), the phase shift $\Phi_{\rm{B,Vn}}$ always decreases with height and with the frequency of the waves. Therefore, at the upper region of the atmosphere, the departure from the value corresponding to the strongly coupled case is larger for $R_{\rm{B,Vn}}$ than for $R_{\rm{B,Vc}}$. The reason is that the neutral fluid only feels indirectly the effect of the magnetic field, by means of the interaction with the charges, and thus, as the collisional coupling is reduced, the neutral component is less affected by the perturbations of the magnetic field. Then, the comparison of the right panels from Figs. \ref{fig:ratio_vels}-\ref{fig:ratio_bvn} confirms that $\Phi_{\rm{B,Vn}} = \Phi_{\rm{B,Vc}} + \Phi_{\rm{V}}$.

    Again, equivalent results are found for backward propagating Alfvén waves but are not presented in this work.

\subsection{Ion-cyclotron and whistler waves} \label{sec:hall}
    In this section, we analyze how the inclusion of Hall's effect in the two-fluid equations modifies the properties of Alfvénic waves and we pay attention to the dissimilarities on the behaviors of the whistler and ion-cyclotron modes. Again, we focus in the case of perturbations generated by a periodic driver (with real frequency but complex wavenumber). Thus, it is convenient to rewrite Eq. (\ref{eq:alfven_dr_hall}) in the following form:
    \begin{equation} \label{eq:kz2_hall}
        k_{z\pm}^{2} = \frac{\omega^{2}}{c_{\rm{A}}^{2}} \frac{\omega + i \left(1 + \chi \right) \nu_{\rm{nc}}}{\omega \left(1 \mp \frac{\omega}{\Omega_{\rm{i}}} \right) + i \nu_{\rm{nc}} \left(1 \mp \left(1 + \chi \right) \frac{\omega}{\Omega_{\rm{i}}} \right)},
    \end{equation}
    where, for simplicity, we have dropped the subscript from $\omega$. 

    Before solving this dispersion relation for a general scenario, we can obtain useful information about the characteristic properties of the two circular polarizations of waves by considering the particular cases of uncoupled and strongly coupled fluids. For the former, if we impose $\nu_{\rm{nc}} = 0$ in Eq. (\ref{eq:kz2_hall}), we get that
    \begin{equation} \label{eq:kz2_hall_nocoll}
        k_{z\pm}^{2} = \frac{\omega^{2}}{c_{\rm{A}}^{2} \left(1 \mp \frac{\omega}{\Omega_{\rm{i}}}\right)},
    \end{equation}
    which corresponds to the dispersion relation for fully ionized plasmas in the single-fluid approximation \citep{Lighthill1960RSPTA.252..397L,Cramer2001paw..book.....C,Huba2003LNP...615..166H}. This expression reveals a fundamental difference between the left-handed and right-handed modes: as the frequency of the driver approaches the cyclotron frequency of the ions ($\Omega_{\rm{i}}$), the wavenumber of the ion-cyclotron modes ($L$) grows without bound ($k_{z+}^{2} \to \infty$), while the wavenumber of the whistler modes ($R$) remains finite. At the singularity $\omega = \Omega_{\rm{i}}$, commonly known as ion cyclotron resonance, the phase speed of the $L$ mode is zero, which means that the perturbation does not propagate. For wave frequencies larger than the cyclotron frequency, the wavenumber of the left-handed polarization becomes a purely imaginary quantity: there is a cutoff region where these waves are evanescent. The right-hand polarized waves are not affected by either the resonance or the cutoff region. We note that here we have assumed again that $\omega > 0$; if we consider negative values of the frequencies, the previously described behaviors of the $L$ and $R$ modes are swapped (that is, $k_{z-}^{2} \to \infty$ when $\omega = - \Omega_{\rm{i}}$).

    To obtain the corresponding expression for the perfect coupling regime, we impose that $\nu_{\rm{nc}} \to \infty$ in Eq. (\ref{eq:kz2_hall}), which leads to
    \begin{equation} \label{eq:kz2_hall_strong}
        k_{z\pm}^{2} = \frac{\omega^{2}}{\frac{c_{\rm{A}}^{2}}{1+\chi} \left(1 \mp \frac{\omega}{\Omega_{\rm{H}}}\right)},
    \end{equation}
    where $\Omega_{\rm{H}}$ is the Hall frequency already defined in Eq. (\ref{eq:omega_hall}). The comparison between Eqs. (\ref{eq:kz2_hall_nocoll}) and (\ref{eq:kz2_hall_strong}) shows that the modified Alfvén speed plays in weakly ionized plasmas the same role as the classical Alfvén speed in fully ionized plasmas and that the existence of resonances and cutoff regions is related to the Hall frequency instead of directly to the cyclotron frequency \citep{Pandey2008MNRAS.385.2269P,Pandey2015MNRAS.447.3604P}. Since in a weakly ionized plasma $\Omega_{\rm{H}} \ll \Omega_{\rm{i}}$, the properties of the ion-cyclotron and the whistler waves are expected to start to differ at a lower range of frequencies than in the case of fully ionized plasmas.

    In the remaining of this section we consider the general case of the solutions to the dispersion relation applying the physical conditions described in Section \ref{sec:background}.

\subsubsection{Wavenumbers and damping rates}
    \begin{figure*}
        \centering
        \includegraphics[width=0.49\hsize]{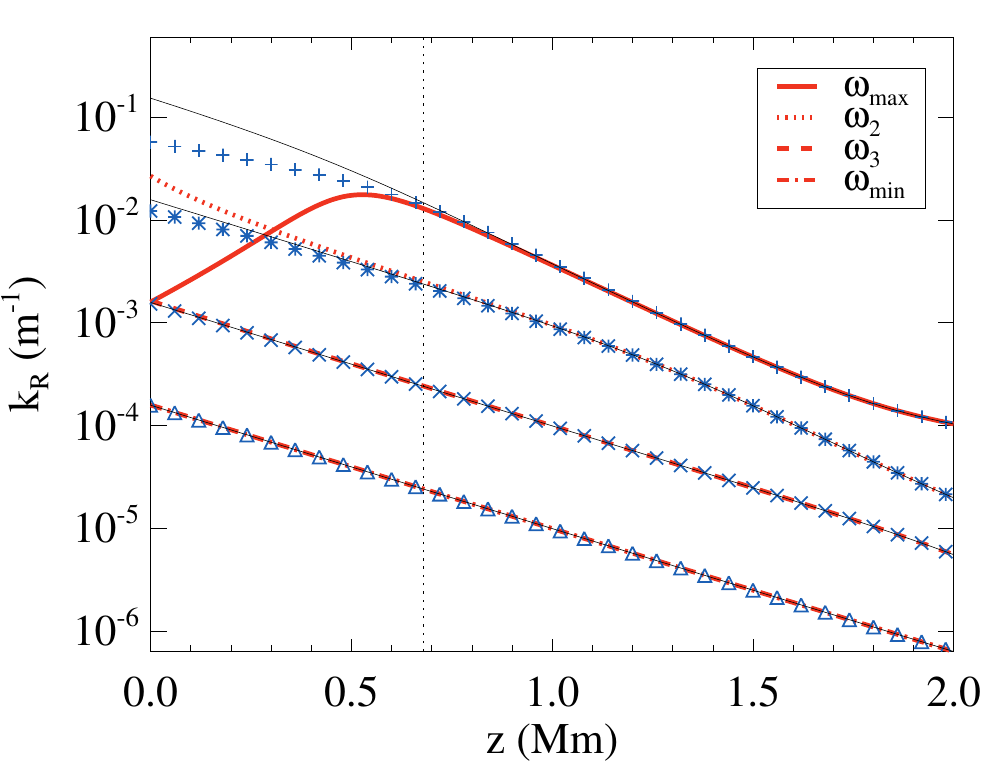}
        \includegraphics[width=0.49\hsize]{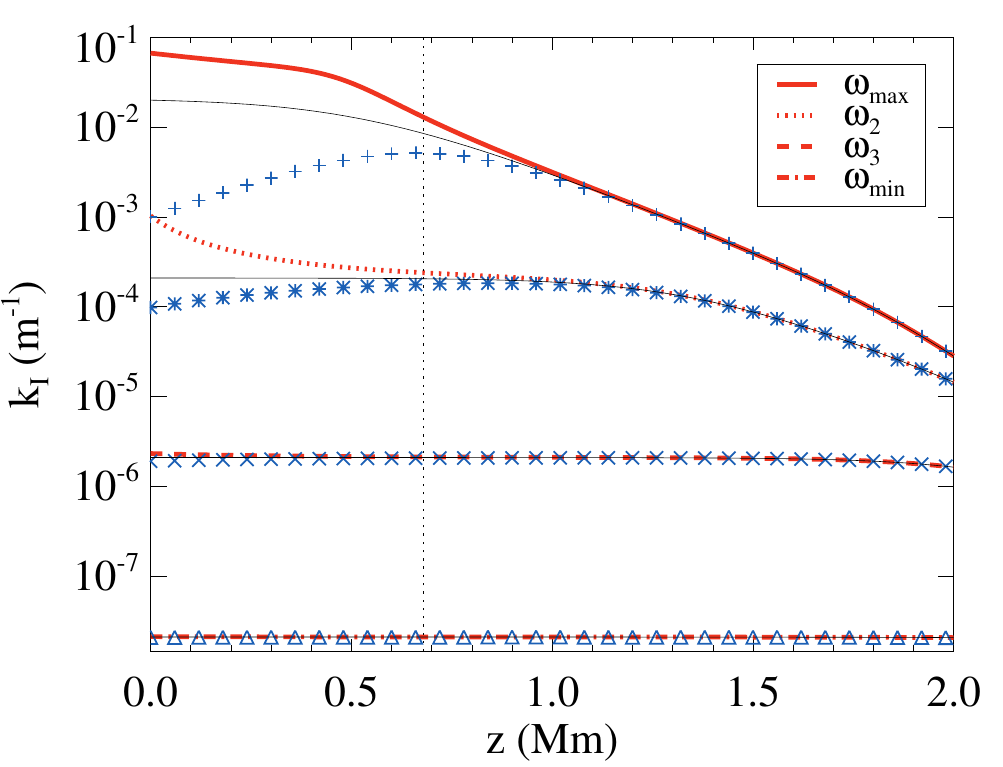}
        \caption{Real part (left panel) and imaginary part (right panel) of the wavenumber of ion-cyclotron (red lines) and whistler (blue symbols) waves. For reference, the solutions for Alfvén waves are included as thin black lines. The vertical dotted line shows the height where $\omega_{\rm{max}} = \Omega_{\rm{H}}$.}
        \label{fig:kRkI_hall}
    \end{figure*}

    We show in Fig. \ref{fig:kRkI_hall} the real part (left panel) and the imaginary part (right panel) of the solutions from Eq. (\ref{eq:kz2_hall}) as functions of height for our model atmosphere. Only the positive solutions have been included. In the first place, we see that the results for the ion-cyclotron modes (red lines) and the whistler modes (blue symbols) both are in perfect agreement with the results for Alfvén waves (black thin lines) at the top layers of the atmosphere for all the periods considered and for the lower range of wave frequencies in the whole atmosphere. This is the expected behavior when $\omega \ll \Omega_{\rm{H}}$. Thus, for the remaining of the paper we will focus only on the bottom layers of the atmosphere, where clear differences are found between the three kinds of waves (the two circularly polarized modes and the linearly polarized Alfvén waves).

    One of the most interesting features depicted on the left panel of Fig. \ref{fig:kRkI_hall} is the total absence of resonances or cutoff regions, that is, regions where $k_{\rm{R}} \to \infty$ or $k_{\rm{R}} = 0$, respectively. From the analysis of the dispersion relation for the strong coupling regime, Eq. (\ref{eq:kz2_hall_strong}), we could expect the existence of a resonant point at $\omega_{\rm{max}} = \Omega_{\rm{H}}$ (marked in the figure as a dotted vertical line) and a cutoff region with $k_{\rm{R}} = 0$ in the lower layers of the atmosphere, where $\omega_{\rm{max}} > \Omega_{\rm{H}}$. However, we see that the wavenumber remains finite and distinct from zero in the whole atmosphere and for both circular polarizations. This is a consequence of taking into account the collisions between the charged and neutral components of the plasma: as it can be checked from Eq. (\ref{eq:kz2_hall}), the inclusion of the terms proportional to $\nu_{\rm{nc}}$ removes the singular point in the denominator and prevents $k_{z\pm}$ from being purely imaginary, in good agreement with the results from \citet{Cramer1997PASA...14..170C} and \citet{MartinezGomez2017ApJ...837...80M}. The same effect has been found analytically and experimentally in fully ionized plasmas for the collisional interaction between charged species \citep[see, e.g.,][]{Cramer2001paw..book.....C,Rahbarnia2010PhPl...17c2102R}. In addition, we see that when $\omega \lesssim \Omega_{\rm{H}}$, the wavenumber of the ion-cyclotron mode is larger than that of the Alfvén wave, while the wavenumber of the whistler mode is smaller. However, when $\omega > \Omega_{\rm{H}}$, both circular polarizations have smaller wavenumbers than the Alfvén mode, since the wavenumber of the ion-cyclotron mode strongly decreases at the deeper layers of the atmosphere.

    On the right panel of Fig. \ref{fig:kRkI_hall} we display the comparison of the damping rates of the three types of waves as functions of height and we see how differently they are affected by the charge-neutral collisional damping. For all the wave frequencies considered in this investigation, we find that the damping rates of the ion-cyclotron modes are larger than those of the Alfvén waves, which are in turn larger than the damping rates of the whistler modes. In addition, the damping rate of whistler modes reaches a maximum as $\omega$ approaches $\Omega_{\rm{H}}$ and then decreases both at the bottom and top layers of the atmosphere, while the value of $k_{\rm{I}}$ keeps increasing for the ion-cyclotron mode as we move to the lower heights. Hence, we see that the attenuation of the perturbations due to charge-neutral collisions strongly depends on the polarization state of the waves, with disparities of up to $2$ orders of magnitude at the bottom of the atmosphere.

\subsubsection{Quality factor and phase speed}
\begin{figure}
        \includegraphics[width=\hsize]{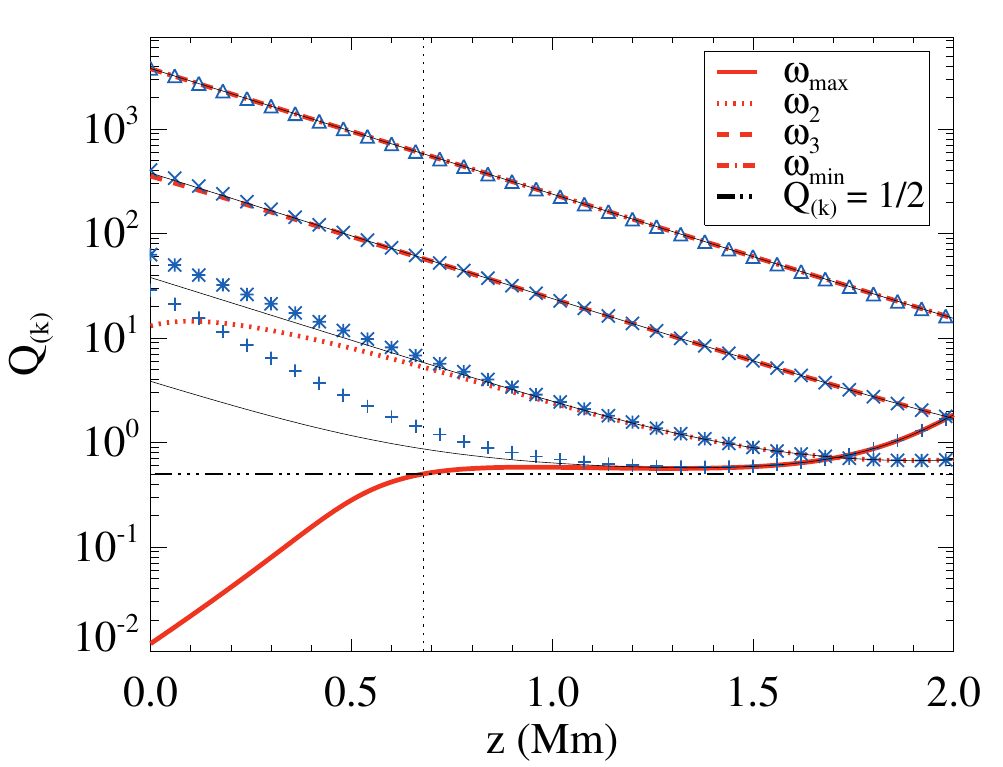}
        \caption{Quality factor of ion-cyclotron (red lines) and whistler (blue symbols) waves as functions of height. The horizontal black line represents the critical value $Q_{(k)} = 1/2$. The results from Fig. \ref{fig:qk_alfven} for Alfvén waves have been included as thin black lines.}
        \label{fig:qk_hall}
    \end{figure}

    Here, we use again Eq. (\ref{eq:q_k}) to compute the quality factor of the solutions analyzed in the previous section. The results are displayed in Fig. \ref{fig:qk_hall}. In general we see similar results to those presented in Fig. \ref{fig:qk_alfven}: the waves are underdamped and $Q_{(k)}$ tends to decrease with height (except for the case of $\omega_{\rm{max}}$). In addition, at the bottom layers the quality factor of the whistler waves is larger than that of Alfvén waves and the quality factor of the cyclotron waves is smaller. However, for the highest-frequency ion-cyclotron wave we find that $Q_{(k)} < 1/2$ at lower heights and the transition from the overdamped to the underdamped regime takes place exactly at the height where $\omega_{\rm{max}} = \Omega_{\rm{H}}$. The small value of $Q_{(k)}$ at the bottom of the atmosphere is related to the strong decrease of the wavenumber and the increase of the damping rate shown in Fig. \ref{fig:kRkI_hall}. Therefore, there is not a strict resonance or cutoff area for the $L$ modes at the lower region of the atmosphere but those waves are still strongly damped in very short distances and the perturbations are not allowed to propagate far away from their source. 
    
\begin{figure}
        \includegraphics[width=\hsize]{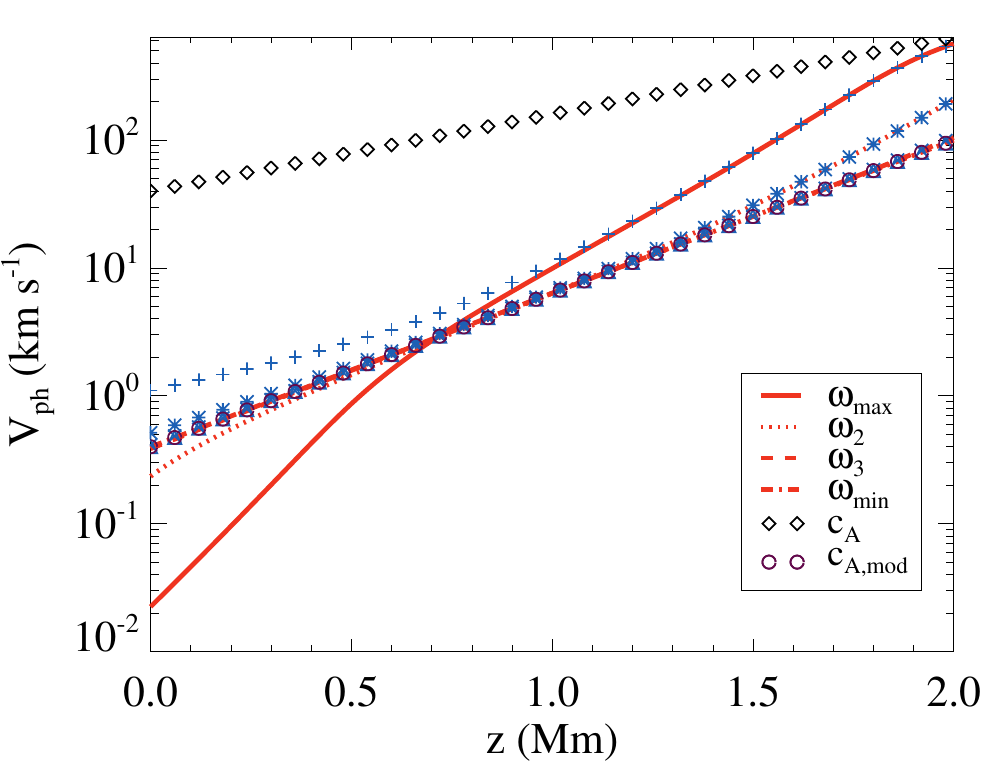}
        \caption{Real part of the phase speeds of ion-cyclotron (red lines) and whistler (blue symbols) waves as functions of height. Diamonds and circles represent the classical and modified versions of the Alfvén speed, $c_{\rm{A}}$ and $c_{\rm{A,mod}}$, respectively.}
        \label{fig:vph_hall}
    \end{figure}

    Then, we use the relation given by Eq. (\ref{eq:vphase2}) to compute the phase speed of the waves. The results are displayed on Fig. \ref{fig:vph_hall}, where the classical and modified Alfvén speeds have also been included for reference. Focusing only on the lower regions of the atmosphere, we see that the phase speed of the right-handed modes is larger than the modified Alfvén speed, while the ion-cyclotron waves propagate at a smaller speed, which strongly decreases when $\omega > \Omega_{\rm{H}}$. These results are consistent with the behavior of the circularly polarized modes in fully ionized plasmas when their propagation speeds are compared to the classical Alfvén speed \citep{Cramer2001paw..book.....C}.
    
\subsubsection{Eigenfunction relations} \label{sec:eigen_hall}
    Now, we come back to Eqs. (\ref{eq:vn_pm})-(\ref{eq:b1_pm}) to compute the relations between the velocities and the magnetic field perturbations in the case of the whistler and ion-cyclotron modes and to check how they compare to the analogous relations for Alfvén waves.

    The first thing we can notice in this study is that from Eq. (\ref{eq:vn_pm}) we get the same velocity ratio than for the case of Alfvén waves, which is given by Eq. (\ref{eq:r_vcvn}) and has already been represented in Fig. \ref{fig:ratio_vels}. This ratio does not depend on the state of polarization of the waves or on their wavenumber, and it is only a function of the frequency of the driver and of the collisional frequency. However, we remark that this result is only applicable to the case of propagating waves generated by a periodic driver. In the case of standing or propagating waves generated by an impulsive perturbation (in which $k_{z}$ would be a real quantity while $\omega$ would be allowed to have both a real and a imaginary part), the frequency is a function of the chosen wavenumber and of the polarization state, so $\omega_{+} \ne \omega_{-}$ \citep{Lighthill1960RSPTA.252..397L,Stix1992wapl.book.....S,Cramer2001paw..book.....C} and the ratios of velocities for the $L$ and $R$ modes are no longer expected to agree with that for Alfvén waves. It would be interesting to explore such scenario in future works.

    On the other hand, we expect that differences appear in the ratios of velocities and magnetic field perturbations, since now we take into account the last term of Eq. (\ref{eq:b1_pm}) that was neglected in Section \ref{sec:eigen_alfvén}. Therefore, the corresponding non-dimensional ratio for whistler and ion-cyclotron waves is given by
    \begin{equation} \label{eq:r_bvc_pm}
        R_{\rm{B,Vc}}^{\pm} = \frac{-k_{z\pm} c_{\rm{A,mod}}}{\omega \pm \frac{k_{z\pm}^{2} c_{\rm{A}}^{2}}{\Omega_{\rm{i}}}},
    \end{equation}
    where $k_{z\pm}$ are the solutions from Eq. (\ref{eq:kz2_hall}). From this expression we can already discuss the most important feature that distinguishes the $L$ and $R$ modes: as explained in the beginning of Section \ref{sec:hall}, at the singular points ($\omega = \Omega_{\rm{i}}$ for a fully ionized plasma and $\omega = \Omega_{\rm{H}}$ for a weakly ionized plasma with strong collisional coupling), the wavenumber of the left-handed modes fulfills that $k_{z+} \to \infty$, while $k_{z-}$ remains finite. Therefore, from Eq. (\ref{eq:r_bvc_pm}) we would obtain that $R_{\rm{B,Vc}}^{+} \to 0$ but $R_{\rm{B,Vc}}^{-} \ne 0$.
    
    More precisely, if we insert Eq. (\ref{eq:kz2_hall_strong}) into Eq. (\ref{eq:r_bvc_pm}) we get the expression for the ratio in the perfect coupling case, namely
    \begin{equation} \label{eq:r_bvc_pm_strong}
        R_{\rm{B,Vc}}^{\pm} = -\sqrt{1 \mp \frac{\omega}{\Omega_{\rm{H}}}},
    \end{equation}
    where, for simplicity, we have only considered the solution with $k_{z} > 0$, which corresponds to forward propagating waves (the expression for backward propagating waves has the same absolute value but it is positive). As expected, in the limit $\omega \ll \Omega_{\rm{H}}$ we recover the appropriate result for Alfvén waves, that is $R_{\rm{B,Vc}} \approx \mp 1$. Then, at the resonance frequency we get that $R_{\rm{B,Vc}}^{+} = 0$ and $R_{\rm{B,Vc}}^{-} = -\sqrt{2}$. The result for the ion-cyclotron mode means that the normalized amplitude of the velocity ($V_{\rm{c+}}/c_{\rm{A,mod}}$) is much larger than that of the magnetic field ($B_{\rm{1,+}}/B_{\rm{0}}$), which is related to the fact that at a resonance the energy of the driver is not transported away but used in increasing the amplitude of the velocity perturbations \citep[see, e.g.,][]{Cramer2001paw..book.....C}.

    For a general value of the wave frequency $\omega > 0$, Eq. (\ref{eq:r_bvc_pm_strong}) shows that $R_{\rm{B,Vc}}^{-} < -1$. This means that for the whistler modes the velocity and magnetic field perturbations of forward propagating waves are in anti-phase ($\Phi_{\rm{B,Vc}}^{-} = \pi$) and that the normalized amplitude of the velocity ($V_{\rm{c-}}/c_{\rm{A,mod}}$) is smaller than the normalized amplitude of the magnetic field ($B_{\rm{1-}}/B_{\rm{0}}$). For backward propagating waves, the perturbations will be in phase ($\Phi_{\rm{B,Vc}}^{-} = 0$) but the relation between their amplitudes would be the same.
    
    In the analysis of the ion-cyclotron modes we have to separate the regime with $\omega < \Omega_{\rm{H}}$ from that with $\omega > \Omega_{\rm{H}}$. In the former, we obtain that $-1 < R_{\rm{B,Vc}}^{+} < 0$, meaning that the perturbations are also in anti-phase but now the velocity has a larger normalized amplitude than the magnetic field. Then, if $\omega > \Omega_{\rm{H}}$, the ratio $R_{\rm{B,Vc}}^{+}$ becomes a pure imaginary quantity and the perturbations are no longer in anti-phase but have a phase shift given by $\Phi_{\rm{B,Vc}}^{+} = -\pi /2$. In addition, in the range $\omega > 2\Omega_{\rm{H}}$, the relation between the amplitudes of the perturbations changes again and now the magnetic field has a larger normalized amplitude than the velocity.

    \begin{figure*}
        \centering
        \includegraphics[width=0.49\hsize]{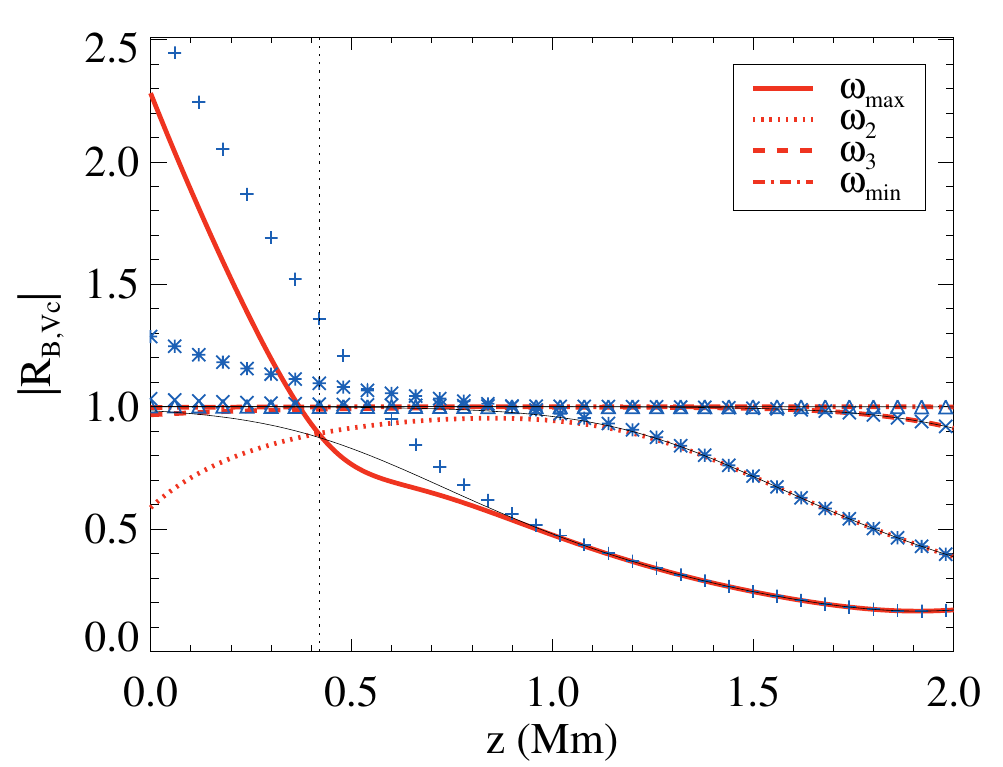}
        \includegraphics[width=0.49\hsize]{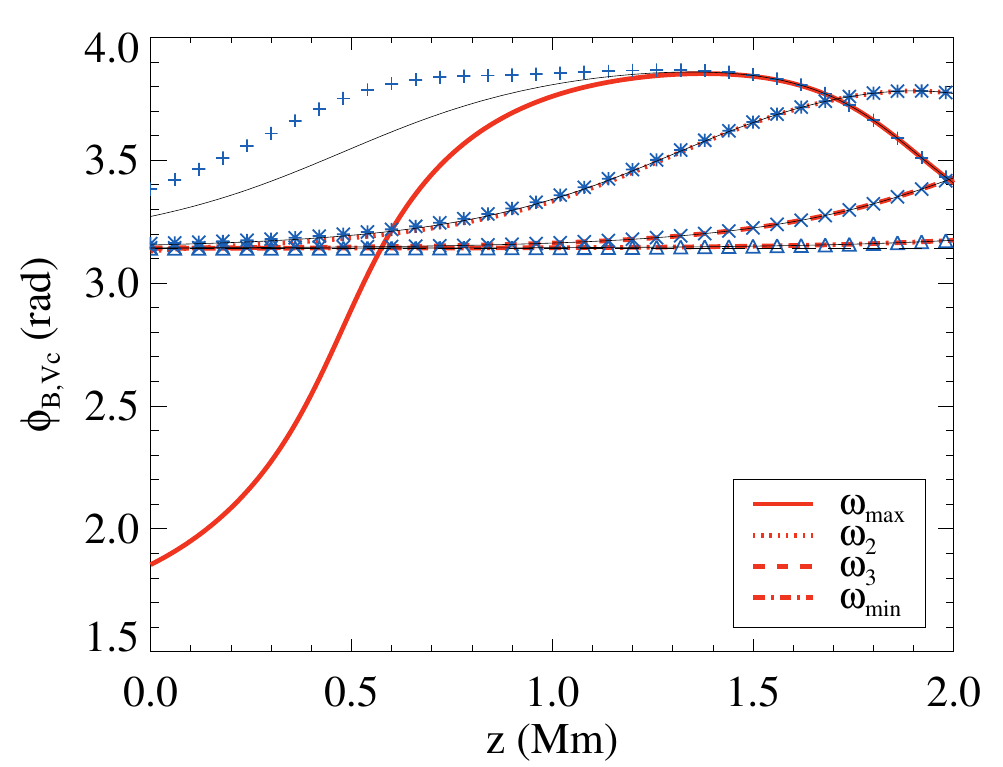}
        \caption{Modulus (left panel) and phases (right panel) of the ratios $R_{\rm{B,Vc}}^{+}$ (red lines) and $R_{\rm{B,Vc}}^{-}$ (blue symbols) as functions of height. The modulus of the ratio, $|R_{\rm{B,Vc}}|$. The vertical dotted line in the left panel shows the position where $\omega_{\rm{max}} = 2 \Omega_{\rm{H}}$. The horizontal black dashed line in the right panel represents the phase angle $\Phi_{\rm{B,Vc}} = \pi$. Results for Alfvén waves have been included as thin black lines.}
        \label{fig:ratio_bvc_hall}
    \end{figure*}

    \begin{figure*}
        \centering
        \includegraphics[width=0.49\hsize]{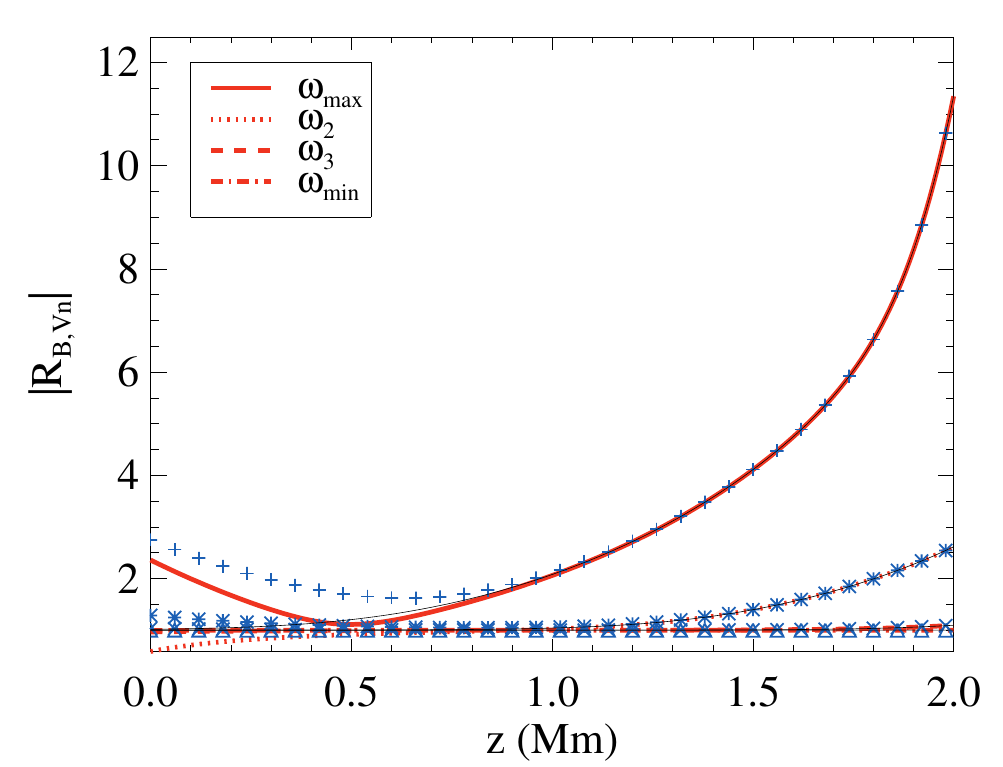}
        \includegraphics[width=0.49\hsize]{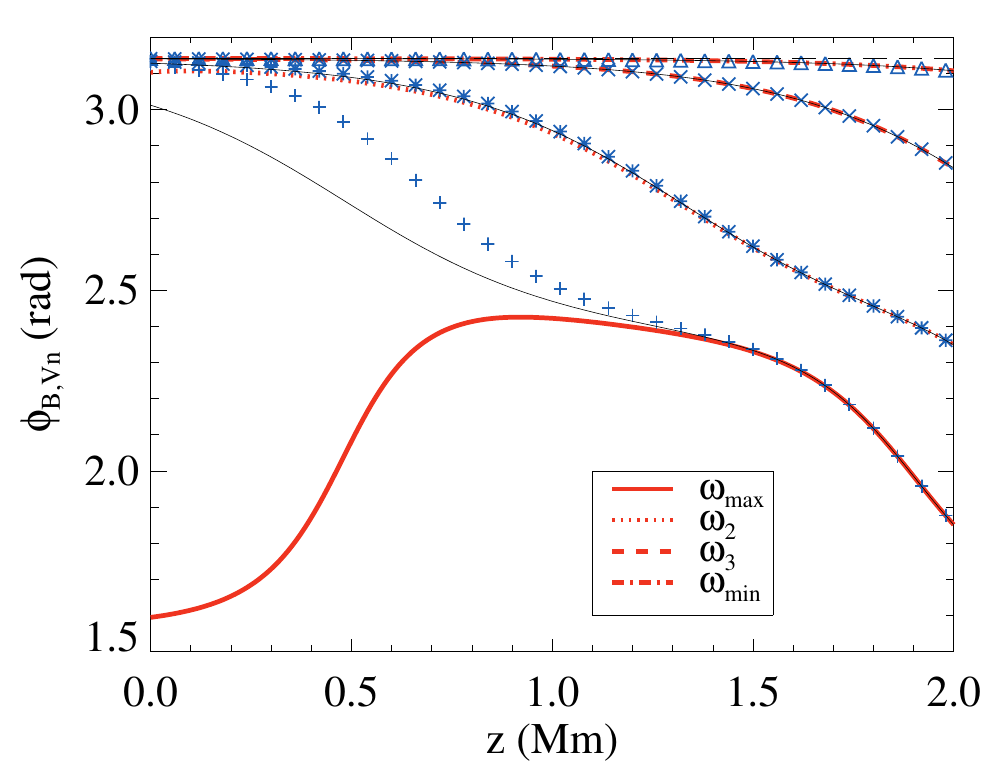}
        \caption{Modulus (left panel) and phases (right panel) of the ratios $R_{\rm{B,Vn}}^{+}$ (red lines) and $R_{\rm{B,Vn}}^{-}$ (blue symbols) as functions of height. The horizontal black dashed line in the right panel represents the phase angle $\Phi_{\rm{B,Vn}} = \pi$. Results for Alfvén waves have been included as thin black lines.}
        \label{fig:ratio_bvn_hall}
    \end{figure*}

    The behavior of the ratios $R_{\rm{B,Vc}}^{\pm}$ detailed above has strong implications regarding the equipartition between the kinetic and the magnetic energies of the wave, as it will be shown in Section \ref{sec:energy}.
    
    After analysing some of the limiting values of the ratios $R_{\rm{B,Vc}}^{\pm}$ we now pay attention to their application to our model atmosphere. The results are presented in Fig. \ref{fig:ratio_bvc_hall}, where their modulus and phases are plotted in the left and the right panels, respectively. We see again that the properties of the $L$ and $R$ only clearly diverge for the bottom half of the atmosphere and for high-frequencies while at the top half we recover the results for Alfvén waves. We find that for a given frequency the modulus and the phase of the ratio for the right-handed modes (blue symbols) are always larger than the corresponding values of the left-handed modes (red lines) and Alfvén waves (black thin lines), so the former have larger normalized amplitudes of the magnetic field perturbations. In addition, in the comparison between the ion-cyclotron and the Alfvén modes we find a change of tendency at the height where $\omega_{\rm{max}} = 2 \Omega_{\rm{H}}$ (marked in this plot by the vertical dotted line): at the bottom layers where $\omega_{\rm{max}} > 2 \Omega_{\rm{H}}$, the value of $|R_{\rm{B,Vc}}^{+}|$ is larger than the one corresponding to Alfvén waves, but becomes smaller when $\omega < 2 \Omega_{\rm{H}}$. Another remarkable feature of Fig. \ref{fig:ratio_bvc_hall} is the absence of any region where $R_{\rm{B,Vc}}^{+} = 0$: this is a consequence of the collisional friction removing the strict cyclotron resonances.

    To conclude the study of the eigenfunction relations, we pay attention to the ratio between the magnetic field and the velocity of the neutral fluid, $R_{\rm{B,Vn}}$, which can be obtained by multiplying Eqs. (\ref{eq:r_vcvn}) and (\ref{eq:r_bvc_pm}). The modulus and phase of this ratio as functions of height are represented in Fig. \ref{fig:ratio_bvn_hall}. There are two main differences with respect to the results for Alfvén waves displayed in Fig. \ref{fig:ratio_bvn}: 1) in the left panel we find that the modulus of both $R_{\rm{B,Vn}}^{+}$ and $R_{\rm{B,Vn}}^{-}$ are larger than $1$ for the high-frequency waves and lower heights; and 2) the phase shift of the ion-cyclotron mode with $\omega_{\rm{max}}$ tends to $\Phi_{\rm{B,Vn}}^{+} \approx \pi / 2$ at the bottom of the atmosphere instead of approaching the value $\Phi_{\rm{B,Vn}} \approx \pi$ (as it happens for the rest of the considered modes and frequencies). 

\subsection{Wave energy} \label{sec:energy}
    An evolution equation for the energy density of the small-amplitude waves can be obtained by applying the procedure described in, for instance \citet{Braginskii1965RvPP....1..205B}, \citet{Bray1974soch.book.....B}, \citet{Walker2004SPP....16.....W,Walker2014_angeo-32-1495-2014}, \citet{Soler2016A&A...592A..28S,Soler2017ApJ...840...20S}. Thus, we take the dot product of $\bm{V}_{\rm{n}}$, $\bm{V}_{\rm{c}}$, and $\bm{B}_{\rm{1}}/\mu_{\rm{0}}$ with Eqs. (\ref{eq:lin_momn}), (\ref{eq:lin_momc}), and (\ref{eq:lin_induc}), respectively, and neglect the terms related to the pressure gradients (since in the present investigation we are focusing on incompressible perturbations). Adding the resulting expressions we get that the temporal evolution of the wave energy is described by
    \begin{equation} \label{eq:energy_equation}
        \frac{\partial \mathcal{U}}{\partial t} + \nabla \cdot \bm{\Pi} = - Q,
    \end{equation}
    where $\mathcal{U}$ is the total energy density of the wave, $\bm{\Pi}$ is the energy density flux, and the right-hand of the equation is related to the loss of wave energy due to the charge-neutral collisions.

    The total energy density is computed as
    \begin{equation} \label{eq:energies_defs}
        \mathcal{U} = \mathcal{K}_{\rm{c}} + \mathcal{K}_{\rm{n}} + \mathcal{M} = \frac{1}{2}\rho_{\rm{c0}}|\bm{V}_{\rm{c}}|^{2} + \frac{1}{2}\rho_{\rm{n0}}|\bm{V}_{\rm{n}}|^{2} + \frac{1}{2\mu_{\rm{0}}} |\bm{B}_{\rm{1}}|^{2},
    \end{equation}
    where $\mathcal{K}_{\rm{c}}$, $\mathcal{K}_{\rm{n}}$, and $\mathcal{M}$ are the contributions from the kinetic energy of charges, the kinetic energy of neutrals and the magnetic energy, respectively.

    The wave energy flux is in this case identical to the magnetic Poynting flux, and it can be written as
    \begin{eqnarray} \label{eq:energy_flux}
        \bm{\Pi} &=& \bm{\Pi}_{\rm{ideal}} + \bm{\Pi}_{\rm{Hall}} = \frac{1}{\mu_{\rm{0}}} \left[ \left(\bm{B}_{\rm{0}} \cdot \bm{B}_{\rm{1}} \right) \bm{V}_{\rm{c}} - \left(\bm{V}_{\rm{c}} \cdot \bm{B}_{\rm{1}} \right) \bm{B}_{\rm{0}} \right] \nonumber \\
        &+& \frac{1}{e n_{\rm{e0}} \mu_{\rm{0}}} \left[ \left(\bm{B}_{\rm{1}} \cdot \bm{J}_{\rm{1}} \right) \bm{B}_{\rm{0}} - \left(\bm{B}_{\rm{1}} \cdot \bm{B}_{\rm{0}} \right) \bm{J}_{\rm{1}} \right],
    \end{eqnarray}
    with $\bm{J}_{\rm{1}} = \left(\nabla \times \bm{B}_{\rm{1}} \right) / \mu_{\rm{0}}$ the perturbation of the current density.

    Nevertheless, it can be checked that here the term $\bm{\Pi}_{\rm{Hall}}$ vanishes, since we have that $\bm{B}_{\rm{1}} \cdot \bm{J}_{\rm{1}} = \bm{B}_{\rm{1}} \cdot \left(i \bm{k} \times \bm{B}_{\rm{1}} \right) / \mu_{\rm{0}} = 0$ and linear Alfvénic waves fulfill the relation $\bm{B}_{\rm{1}} \cdot \bm{B}_{\rm{0}} = 0$. We note that this would no longer be true when considering non-linear Alfvénic waves due to the fact that they may generate perturbations along the direction longitudinal to $\bm{B}_{\rm{0}}$ \citep[see, e.g.,][]{Hollweg1971JGR....76.5155H,Rankin1994JGR....9921291R}. However, in the present scenario only the ideal term contributes to the Poynting flux, which is then given by
    \begin{equation} \label{eq:poynting_ideal}
        \bm{\Pi} = \bm{\Pi}_{\rm{ideal}} = -\frac{1}{\mu_{\rm{0}}}\left(\bm{V}_{\rm{c}} \cdot \bm{B}_{\rm{1}} \right) \bm{B}_{\rm{0}},
    \end{equation}
    showing that the energy of the Alfvénic waves is propagated along the direction of the background magnetic field. Then, using the relation between the perturbations of magnetic field and velocity of charges given by Eq. (\ref{eq:b1_vc}) it can be checked that $\bm{\Pi} / \bm{B}_{\rm{0}} > 0$ for forward propagating waves and $\bm{\Pi} / \bm{B}_{\rm{0}} < 0$ for backward propagating waves.

    Finally, the quantity $Q$ is given by
    \begin{equation} \label{eq:qterm}
        Q = \alpha_{\rm{cn}} \left(\bm{V}_{\rm{c}} - \bm{V}_{\rm{n}} \right)^{2},
    \end{equation}
    and, if we consider the full non-linear evolution described by Eqs. (\ref{eq:mom_n})-(\ref{eq:indu}) and assume that the friction coefficient $\alpha_{\rm{cn}}$ does not vary with time, it can be identified with the total heating term resulting from the addition of Eqs. (\ref{eq:qn}) and (\ref{eq:qc}), that is $Q = Q_{\rm{nc}} + Q_{\rm{cn}}$ \citep{Soler2016A&A...592A..28S,Soler2017ApJ...840...20S,Soler2019ApJ...871....3S}. Therefore, if the total energy conservation is taken into account, the energy lost by the waves is converted into internal energy of the plasma, leading to an increase in temperature \citep{Braginskii1965RvPP....1..205B,Draine1986MNRAS.220..133D}. We remark, however, that the heating of the plasma cannot be properly studied by the linear analysis performed in this work.

    It is a classical result that there is equipartition between the kinetic energy and the magnetic energy for Alfvén waves in fully ionized plasmas \citep{Walen1944ArMAF..30A...1W,Ferraro1958ApJ...127..459F,Braginskii1965RvPP....1..205B,Priest1984smh..book.....P}. Therefore, it is interesting to check if that relation still holds when the effects of charge-neutral collisions and Hall's term are included. To perform this study it is useful to express the kinetic energy density of the neutral fluid and the magnetic energy density in terms of the kinetic energy density of the charged fluid. Thus, we resort to the eigenfunction relations that can be derived from Eqs. (\ref{eq:vn_pm}) and (\ref{eq:b1_pm}), which have already been analyzed in Sections \ref{sec:eigen_alfvén} and \ref{sec:eigen_hall}. In addition, we take into account that the perturbations are complex quantities, so $|\bm{V}_{\rm{c}}|^{2} = \bm{V}_{\rm{c}} \bm{V}_{\rm{c}}^{*}$, $|\bm{V}_{\rm{n}}|^{2} = \bm{V}_{\rm{n}} \bm{V}_{\rm{n}}^{*}$, and $|\bm{B}_{\rm{1}}|^{2} = \bm{B}_{\rm{1}} \bm{B}_{\rm{1}}^{*}$, where ${}^{*}$ denotes the complex conjugate of the variable.

    First, we study the case of Alfvén waves, neglecting the presence of the Hall term in the induction equation. Thus, using the ratio of velocities given by Eq. (\ref{eq:r_vcvn}) we have that
    \begin{equation} \label{eq:vnvc2}
        |\bm{V}_{\rm{n}}|^{2} = \frac{\nu_{\rm{nc}}^{2}}{\omega^{2} + \nu_{\rm{nc}}^{2}}|\bm{V}_{\rm{c}}|^{2},
    \end{equation}
    and the total kinetic energy density, $\mathcal{K}_{\rm{T}} = \mathcal{K}_{\rm{c}} + \mathcal{K}_{\rm{n}}$ is given by
    \begin{equation} \label{eq:ek1_tot}
        \mathcal{K}_{\rm{T}} = \mathcal{K}_{\rm{c}} \left[1 + \frac{\chi \nu_{\rm{nc}}^{2}}{\omega^{2} + \nu_{\rm{nc}}^{2}} \right] = \mathcal{K}_{\rm{c}} \left[\frac{\omega^{2} + \left(1 + \chi \right) \nu_{\rm{nc}}^{2}}{\omega^{2} + \nu_{\rm{nc}}^{2}} \right],
    \end{equation}
    where the relation $\rho_{\rm{n0}} = \chi \rho_{\rm{c0}}$ has been applied.

    The expression for the magnetic energy density is obtained after inserting Eq. (\ref{eq:alfven_dr2}) into Eq. (\ref{eq:b1_vc}) and computing the product $\bm{B}_{\rm{1}}\bm{B}_{\rm{1}}^{*}/(2 \mu_{\rm{0}})$, leading to
    \begin{equation} \label{eq:eb1_alfvén}
        \mathcal{M}_{\rm{A}} = \mathcal{K}_{\rm{c}} \sqrt{\frac{\omega^{2} + \left(1+\chi\right)^{2}\nu_{\rm{nc}}^{2}}{\omega^{2} + \nu_{\rm{nc}}^{2}}}.
    \end{equation}

    It can be checked that in the limit of no collisional coupling ($\nu_{\rm{nc}} = 0$), the classical equipartition relation, $\mathcal{M}_{\rm{A}} = \mathcal{K}_{\rm{T}} = \mathcal{K}_{\rm{c}}$, is recovered; here, only the charged fluid contributes to the total kinetic energy because the neutral component is not affected by the magnetic field and its velocity is zero. In the opposite limit, when $\nu_{\rm{nc}} \to \infty$, we find that $\mathcal{M}_{\rm{A}} = \mathcal{K}_{\rm{T}} = \mathcal{K}_{\rm{c}} \left(1 + \chi \right)$, meaning that there is again equipartition between the magnetic and the total kinetic energy, but now the latter takes into account the contribution from the neutral fluid too. However, from the inspection of Eqs. (\ref{eq:ek1_tot}) and (\ref{eq:eb1_alfvén}) we do not expect that the relation holds in the intermediate regime of collisional coupling. Before analysing in detail this regime, we move to the more general case in which the effect of Hall's term is included.

    To compute the expressions for the magnetic energy and the total kinetic energy when Hall's term is taken into account, it is convenient to rewrite the dispersion relation given by Eq.
    (\ref{eq:kz2_hall}) as
    \begin{equation} \label{eq:kz2_hall_gamma}
        k_{z\pm}^{2} = \frac{\omega^{2}}{c_{\rm{A}}^{2}}\frac{\omega + i \left(1 + \chi \right) \nu_{\rm{nc}}}{\omega \gamma_{\pm} + i \nu_{\rm{nc}} \eta_{\pm}},
        \end{equation}
    where
    \begin{equation} \label{eq:gamma_delta}
        \gamma_{\pm} = 1 \mp \frac{\omega}{\Omega_{\rm{i}}} \quad \text{and} \quad \eta_{\pm} = 1 \mp \frac{\omega}{\Omega_{\rm{H}}}
    \end{equation}
    are real parameters associated with the cyclotron frequency and the Hall frequency, respectively.

    For propagating waves generated by a periodic driver, the inclusion of Hall's term does not change the expressions that relate the velocity of neutrals with the velocity of charges, and the kinetic energy of the charges with the total kinetic energy. Thus, Eqs. (\ref{eq:vnvc2}) and (\ref{eq:ek1_tot}) still hold (we expect that this would not be applicable for standing or propagating waves generated by an impulsive driver, and it should be studied in the future). In contrast, the expression for the magnetic energy density becomes more convoluted:
    \begin{equation} \label{eq:eb1_hall}
        \mathcal{M}_{\pm} = \mathcal{K}_{\rm{c}} \sqrt{\frac{\omega^{2} + \left(1 + \chi \right)^{2} \nu_{\rm{nc}}^{2}}{\omega^{2} \gamma_{\pm}^{2} + \nu_{\rm{nc}}^{2} \eta_{\pm}^{2}}} \frac1{\Gamma_{\pm}},
   \end{equation}
   with
   \begin{eqnarray} \label{eq:Gamma_pm}
        \Gamma_{\pm} &=& 1 \pm \frac{2 \omega}{\Omega_{\rm{i}}} \left(\frac{\omega^{2} \gamma_{\pm} + \left(1 + \chi \right) \nu_{\rm{nc}}^{2} \eta_{\pm}}{\omega^{2} \gamma_{\pm}^{2} + \nu_{\rm{nc}}^{2} \eta_{\pm}^{2}}\right) \nonumber \\
        &+& \frac{\omega^{2}}{\Omega_{\rm{i}}^{2}} \left(\frac{\omega^{2} + \left(1 + \chi \right)^{2} \nu_{\rm{nc}}^{2}}{\omega^{2} \gamma_{\pm}^{2} + \nu_{\rm{nc}}^{2} \eta_{\pm}^{2}} \right)
    \end{eqnarray}
    It is worth analysing the limits of Eq. (\ref{eq:eb1_hall}). In the low-frequency limit, where $\omega \ll \Omega_{\rm{H}} < \Omega_{\rm{i}}$, we have that $\gamma_{\pm} = \eta_{\pm} = \Gamma_{\pm} = 1$ and we recover the expression of the magnetic energy for Alfvén waves, Eq. (\ref{eq:eb1_alfvén}). Then, in the limit of no collisions between the two fluids ($\nu_{\rm{nc}} = 0$), we find that
    \begin{equation} \label{eq:eb1_hall_nocoll}
        \mathcal{M}_{\pm} = \mathcal{K}_{\rm{c}} \gamma_{\pm} = \mathcal{K}_{\rm{c}} \left(1 \mp \frac{\omega}{\Omega_{\rm{i}}} \right),
    \end{equation}
    which shows that $\mathcal{M}_{+} < \mathcal{K}_{\rm{c}}$ and $\mathcal{M}_{-} > \mathcal{K}_{\rm{c}}$, meaning that the magnetic energy is smaller than the kinetic energy for the ion-cyclotron waves but it is larger for the whistler modes, as already discussed by \citet{Campos1992_10.1063/1.860136}. Therefore, the equipartition relation is not fulfilled as the frequency of the waves approaches the cyclotron frequency \citep[see, e.g.,][]{Galtier2006JPlPh..72..721G,Mahajan2005MNRAS.359L..27M,Lingam2016MNRAS.460..478L,Pouquet2020Atmos..11..203P}. Nevertheless, we also find that the relation
    \begin{equation} \label{eq:sum_eb1_hall}
        \frac{\mathcal{M}_{+} + \mathcal{M}_{-}}{2} = \mathcal{K}_{\rm{c}}
    \end{equation}
    holds for any wave frequency, which might be interpreted as a modified version of the equipartition relation.
    
    A surprising consequence of Eq. (\ref{eq:eb1_hall_nocoll}) is that the magnetic energy density of the ion-cyclotron modes, $\mathcal{M}_{+}$, becomes negative when $\omega > \Omega_{\rm{i}}$ and then the total energy density, $\mathcal{M}_{+} + \mathcal{K}_{\rm{c}}$, becomes negative when $\omega > 2 \Omega_{\rm{i}}$. The study of the implications of a negative total energy of the wave is out of the scope of the present paper. Nevertheless, we remind that in this range of frequencies the ion-cyclotron waves are affected by a cutoff region, so they are evanescent. Thus, this analysis of the wave energy based on normal modes has to be taken with some caution, and a proper investigation would require the consideration of the full temporal evolution of the perturbations.
    
    Applying the strong coupling condition ($\nu_{\rm{nc}} \gg \omega$) to Eq. (\ref{eq:eb1_hall}) we obtain
    \begin{equation} \label{eq:eb1_hall_strong}
        \mathcal{M}_{\pm} = \mathcal{K}_{\rm{c}} \left(1 + \chi \right) \eta_{\pm} = \mathcal{K}_{\rm{T}} \left(1 \mp \frac{\omega}{\Omega_{\rm{H}}} \right),
    \end{equation}
    which would present the same behavior as Eq. (\ref{eq:eb1_hall_nocoll}) but substituting $\mathcal{K}_{\rm{c}}$ by $\mathcal{K}_{\rm{T}}$ and $\Omega_{\rm{i}}$ by $\Omega_{\rm{H}}$, and an analogous relation to Eq. (\ref{eq:sum_eb1_hall}) can be obtained.

    \begin{figure}
        \includegraphics[width=\hsize]{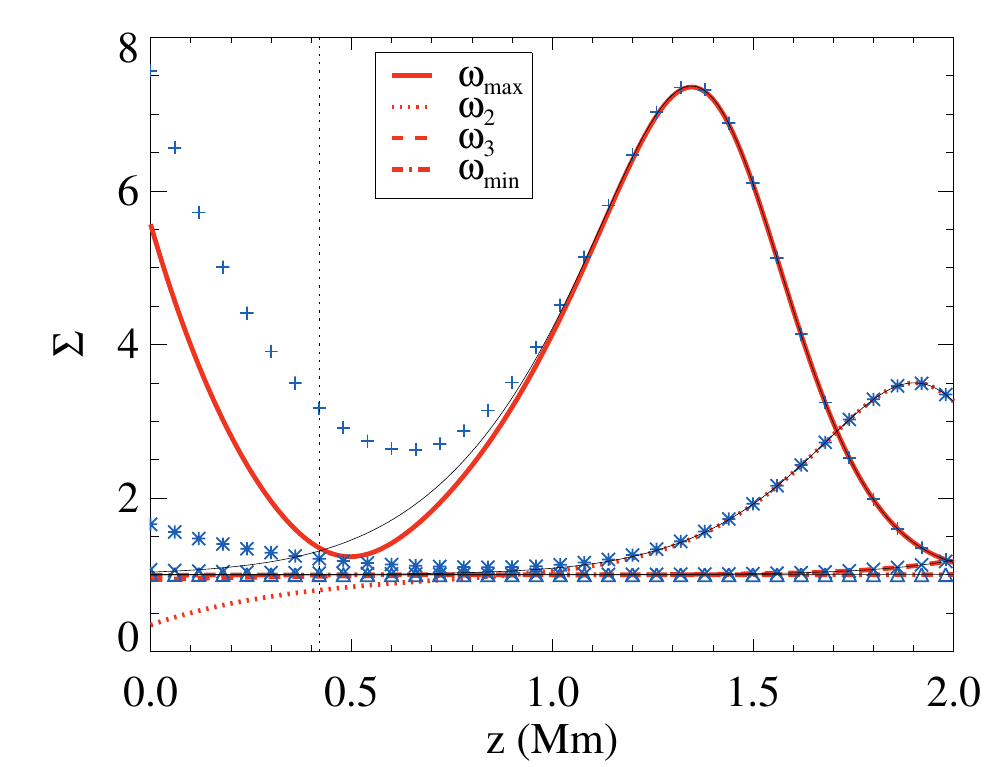}
        \caption{Ratio of magnetic to kinetic energy density as function of height for Alfvén waves (black thin lines), ion-cyclotron modes (red lines) and whistler modes (blue symbols). The dotted vertical line represents the height where $\omega_{\rm{max}} = 2 \Omega_{\rm{H}}$.}
        \label{fig:magkin_ratio}
    \end{figure}

    Now we focus on the full expression of the magnetic energy density, Eq. (\ref{eq:eb1_hall}), and check how an imperfect collisional coupling between the two components of the plasma modifies the results described in the previous paragraphs. We present in Fig. \ref{fig:magkin_ratio} the ratios of magnetic energy to total kinetic energy ($\Sigma = \mathcal{M} / \mathcal{K}_{\rm{T}}$) as functions of height for the Alfvén (black thin lines), ion-cyclotron (red lines) and whistler (blue symbols) modes, which will be denoted as $\Sigma_{\rm{A}}$, $\Sigma_{+}$, and $\Sigma_{-}$, respectively. The results for the three kinds of modes collapse to the value $\Sigma = 1$ in the regime of low frequency waves, recovering the classical equipartition between the magnetic and the total kinetic energy densities \citep{Walen1944ArMAF..30A...1W,Priest1984smh..book.....P}. However, as the frequency of the driving perturbation is increased, the equipartition relation is not fulfilled anymore: at the top layers of the atmosphere we see that $\Sigma > 1$ for all the modes, meaning that the magnetic energy density is larger than the kinetic energy density, specially for the wave with $\omega_{\rm{max}}$, which reaches a value of $\Sigma \approx 8$ around $z \approx 1.35 \ \rm{Mm}$.

    As we move towards the bottom of the atmosphere, we see in Fig. \ref{fig:magkin_ratio} that if $\omega < \Omega_{\rm{H}}$ the energy ratios are ordered as $\Sigma_{-} > \Sigma_{\rm{A}} > \Sigma_{\rm{+}}$. This ordering is broken in the region where $\omega_{\rm{max}} > 2 \Omega_{\rm{H}}$ (in the area located to the left of the vertical dotted line): in this case we find that the ratio of both circular polarizations is larger than that for Alfvén waves. Finally, we remark that $\Sigma_{+}$ is always positive in the whole atmosphere and for all the wave frequencies studied here. This means that, due to the effect of the charge-neutral collisions, the magnetic energy of the ion-cyclotron modes can be smaller than their total kinetic energy but it is never negative in these conditions and, consequently, the total wave energy remains always positive.

    In summary, we find that the classical equipartition relation for Alfvén waves in fully ionized plasmas holds for weakly ionized plasmas in the limits of strong and weak collisional coupling, $\omega \ll \nu_{\rm{nc}}$ and $\omega \gg \nu_{\rm{nc}}$, respectively \citep[see, e.g.,][]{Zweibel1995ApJ...439..779Z}. However, it breaks when $\omega \sim \nu_{\rm{nc}}$ (due to the interaction of the two species), and when $\omega \sim \Omega_{\rm{H}}$ (due to the effect of Hall's term). The dominance of the magnetic energy over the kinetic energy related to the collisional interaction may be explained by the kinetic energy of the waves being more efficiently converted into internal energy of the plasma than their magnetic energy. However, a better understanding of this matter requires the use of a non-linear approach that fully takes into account the heating of the plasma.
      
\section{Summary and discussion} \label{sec:concl}
    In the present paper we have performed a comprehensive study of the properties of small-amplitude Alfvénic waves propagating in weakly ionized plasmas. We have used a two-fluid model in which one of the fluids contains all the electrically charged species of the plasma and the other fluid is composed by the neutral species \citep[see, e.g.,][]{Khomenko2014PhPl...21i2901K}. In this model, the two fluids are allowed to interact by means of elastic collisions. In addition, the effect of Hall's current has also been taken into account.
    
    For the sake of simplicity, we have considered that the plasma is only composed of hydrogen and we have used as a reference a toy model of the solar chromosphere, which is described as an isothermal vertically stratified atmosphere embedded in a uniform magnetic field. However, as we have shown, the properties of the studied waves mainly depend on the relation between their oscillation frequencies and the characteristic frequencies of the system, such as those related to the charge-neutral collisions ($\nu_{\rm{nc}}$ and $\nu_{\rm{cn}}$) and those related to the cyclotron motions of the plasma ($\Omega_{\rm{i}}$ and $\Omega_{\rm{H}}$). Therefore, we expect that the results discussed here can be easily generalized to other astrophysical scenarios with different plasma compositions like planetary ionospheres, the interstellar medium, or molecular clouds.

    In the first place, we have applied a normal mode analysis to the linearized version of the two-fluid evolution equations in order to derive the dispersion relation for incompressible and transverse small-amplitude perturbations propagating along the background magnetic field. In the range of low frequencies ($\omega \ll \Omega_{\rm{H}}, \Omega_{\rm{i}}$) we have recovered the dispersion relation that has already been derived by, for instance, \citet{Kulsrud1969ApJ...156..445K}, \citet{Kumar2003SoPh..214..241K}, \citet{Zaqarashvili2011A&A...529A..82Z}, \citet{Mouschovias2011MNRAS.415.1751M} or \citet{Soler2013ApJ...767..171S}, which describes the properties of linearly polarized Alfvén waves in partially ionized plasmas. We have studied how the wavenumber of the perturbations ($k_{\rm{R}}$) and the damping rates due to charge-neutral collisions ($k_{\rm{I}}$) depend on the relation between the wave frequency, $\omega$, and the collision frequency, $\nu_{\rm{nc}}$, confirming various already known results. For instance, in the strong coupling conditions ($\nu_{\rm{nc}} \gg \omega$) found at the lower layers of our model atmosphere, the damping rates are proportional to the square of the wave frequency (so high-frequency waves are much more efficiently damped than the low-frequency ones) and the wavenumbers are inversely proportional to the modified Alfvén speed (which takes into account the total density of the plasma). On the other hand, if the wave frequency approaches or becomes larger than the collision frequency, the damping rates become independent from $\omega$ and the wavenumber is inversely proportional to the classical Alfvén speed (which only considers the density of the charged fluid). In addition, we have checked the validity of the analytical approximations to the solutions of the dispersion relation provided by \citet{Soler2013ApJ...767..171S}, finding that they are in very good agreement with the exact solutions in the limits of strong and weak coupling.

    As the frequency of the waves is increased, Hall's current starts to play a fundamental role \citep{Lighthill1960RSPTA.252..397L,Cramer2001paw..book.....C}, specially when the weak ionization of the plasma is taken into account \citep{Amagishi1993PhRvL..71..360A,Pandey2008MNRAS.385.2269P}. In this scenario, the waves are no longer linearly polarized but circularly polarized and their wavenumbers and damping rates strongly depend on their polarization state. As shown in Figs. \ref{fig:kRkI_hall} and \ref{fig:qk_hall}, the left-handed circularly polarized (ion-cyclotron) modes are more efficiently damped by charge-neutral collisions than the linearly polarized Alfvén waves and the right-handed circularly polarized (whistler) modes. Furthermore, the collisional interaction removes the resonances and the cutoff regions of the ion-cyclotron modes that would appear at $\omega \geq \Omega_{\rm{i}}$ (for a fully ionized plasma) and $\omega \geq \Omega_{\rm{H}}$ (for a weakly ionized plasma) if the collisions were not taken into account, in agreement with the findings from \citet{Cramer1997PASA...14..170C}, \citet{Rahbarnia2010PhPl...17c2102R} and \citet{MartinezGomez2017ApJ...837...80M}. Hence, ion-cyclotron waves are no longer evanescent at this range of frequencies, although they are still strongly damped by collisions.

    Apart from the wavenumbers and damping rates directly obtained from the dispersion relations, we have also focused on other properties of the eigenmodes of the two-fluid system, such as their phase speeds and their polarization relations. Regarding the former, we confirm that the propagation speed of the linearly polarized modes varies from the modified Alfvén speed ($c_{\rm{A,mod}}$) in the strong coupling limit \citep[see, e.g.,][]{Kumar2003SoPh..214..241K,Zaqarashvili2011A&A...529A..82Z,Soler2013ApJ...767..171S} to the classical Alfvén speed ($c_{\rm{A}}$) in the weak coupling regime. Then, the whistler modes propagate at faster speeds than the Alfvén waves, and the ion-cyclotron waves propagate at slower speeds \citep[see, e.g.,][]{Campos1992_10.1063/1.860136,Cramer2001paw..book.....C}.
    
    Regarding the polarization relations, we show that a phase shift of up to $|\Phi_{\rm{V}}| = \pi/2$ appears between the velocities of charges and neutrals as the frequency $\omega$ approaches the collision frequency $\nu_{\rm{nc}}$, in good agreement with the numerical results from \citet{MartinezGomez2017ApJ...837...80M,MartinezGomez2018ApJ...856...16M}, \citet{Popescu2019A&A...627A..25P} and \citet{MartinezSykora2020ApJ...900..101M}. Furthermore, we find that the Walén relation for fully ionized plasmas, which states that the perturbations of velocity and magnetic field are in phase for backward propagating Alfvén waves and in anti-phase for the case of forward propagation \citep{Walen1944ArMAF..30A...1W,Priest1984smh..book.....P}, does not hold for high-frequency waves in partially ionized plasmas. The phase shift thus becomes a function of the relation between $\omega$ and $\nu_{\rm{nc}}$. This result may be relevant for the field of observations of oscillations in astrophysical plasmas, since the Walén relation has been commonly used to identify the presence of Alfvén waves in environments such as the solar wind \citep[see, e.g.,][]{Belcher1969JGR....74.2302B,Belcher1971JGR....76.3534B,Burlaga1971SSRv...12..600B}. Now, the existence of phase shifts between the velocity of neutrals, the velocity of charges and the perturbations of magnetic field might be used to identify the presence of high-frequency Alfvénic waves in partially ionized plasmas.

    From the study of the eigenfunction relations we find another formula that might be useful for the interpretation of observational data of partially ionized plasmas. Equation (\ref{eq:r_vcvn}) shows that (for propagating Alfvénic waves) the ratio of velocities of charges and neutrals only depends on the relation between the wave frequency and the collision frequency. Hence, measures of the velocities of the two fluids could be used to estimate the coupling degree of the plasma. Since $\bm{V}_{\rm{c}}$ and $\bm{V}_{\rm{n}}$ are complex quantities, it is more convenient for this computation to use the ratio given by Eq. (\ref{eq:vnvc2}), which can be rewritten as
    \begin{equation} \label{eq:w_nunc}
        \frac{\omega}{\nu_{\rm{nc}}} = \sqrt{\frac{|\bm{V}_{\rm{c}}|^{2}}{|\bm{V}_{\rm{n}}|^{2}} -1}.
    \end{equation}
    In this expression, the only unknown quantity is the collision frequency, $\nu_{\rm{nc}}$, since $\omega$, $\bm{V}_{\rm{c}}$, and $\bm{V}_{\rm{n}}$ can be obtained from observations. For instance, the works of \citet{Khomenko2016ApJ...823..132K}, \citet{Anan2017A&A...601A.103A}, \citet{Wiehr2019ApJ...873..125W,Wiehr2021ApJ...920...47W}, \citet{Stellmacher2017SoPh..292...83S}, \citet{Zapior2022ApJ...934...16Z} and  \citet{Gonzalez2024A&A...681A.114G} have measured drift velocities between ions and neutrals of up to $\sim 1 \ \rm{km \ s^{-1}}$ in solar prominences, which demonstrates that in some regions of these solar structures the different components of the plasma are not perfectly coupled. Therefore, Eq. (\ref{eq:w_nunc}) might be applied to the data from the previously mentioned works (or from similar investigations) to estimate the collisional coupling between the charged and neutral components of the prominence plasma (and other astrophysical environments), under the assumption that the observed oscillations are related to Alfvénic waves.

    In addition, we have investigated the relations between the different components of the wave energy density and their dependence on the frequency of the waves and their polarization state. We have found that the classical equipartition relation between kinetic and magnetic energy densities for Alfvén waves \citep{Walen1944ArMAF..30A...1W,Ferraro1958ApJ...127..459F,Braginskii1965RvPP....1..205B,Priest1984smh..book.....P} does not generally hold when the effects of Hall's current and charge-neutral collisions are addressed. The relation is still fulfilled in the limits of strong collisional coupling ($\nu_{\rm{nc}} \gg \omega$) and of low frequencies ($\omega \ll \Omega_{\rm{H}}$). However, as illustrated in Fig. \ref{fig:magkin_ratio}, we find that the magnetic energy density is larger than the total kinetic energy density for Alfvén waves ($\Sigma_{\rm{A}} > 1$) and, specially, for the whistler modes ($\Sigma_{-} > 1$). The ratio of magnetic to kinetic energy density of the ion-cyclotron modes ($\Sigma_{+}$) strongly depends on the range of frequencies considered: for strong collisional coupling and low wave frequency in comparison with the Hall frequency ($\omega \ll \Omega_{\rm{H}}$), the magnetic energy of these waves is smaller than the kinetic energy ($\Sigma_{+} < 1$); for larger frequencies with $\omega < 2 \Omega_{\rm{H}}$, the magnetic energy is larger than the kinetic energy but with a smaller energy ratio than for Alfvén waves ($1 < \Sigma_{+} < \Sigma_{\rm{A}}$); finally, when $\omega > 2 \Omega_{\rm{H}}$, the energy ratio of ion-cyclotron modes becomes larger than the energy ratio of Alfvén waves ($\Sigma_{+} > \Sigma_{\rm{A}}$), but is still smaller than that for the whistler modes ($\Sigma_{+} < \Sigma_{-}$).

    The lack of equipartition between the magnetic and the kinetic energies has already been reported, for instance, in observations and numerical simulations of Alfvénic turbulence in the solar wind \citep{Belcher1969JGR....74.2302B,Tanenbaum1962PhFl....5.1226T,Dastgeer2000PhPl....7..571D,Boldyrev2012AIPC.1436...18B,Wang2011ApJ...740L..36W}. Usually an excess of magnetic energy has been found \citet{Belcher1971JGR....76.3534B,Matthaeus1982JGR....87.6011M,Bruno1985JGR....90.4373B,Chen2013ApJ...770..125C}. This excess has been associated with the influence of Hall's term at short scales \citep[see, e.g.,][]{Campos1992_10.1063/1.860136,Galtier2006JPlPh..72..721G,Lingam2016MNRAS.460..478L,Lingam2016ApJ...829...51L,Pouquet2020Atmos..11..203P} and, more specifically, with the presence of whistler waves in the fully ionized plasma. The present study supports this interpretation but also extends the investigation to scenarios in which neutrals are involved. Hence, the results discussed here may be of particular relevance for the research of turbulence in partially ionized plasmas \citep[see, e.g,][]{Xu2017NJPh...19f5005X,Xu2017ApJ...850..126X,Benavides2020JFM...900A..28B,Hu2024MNRAS.527.3945H}. 

    Moreover, we have shown that it is fundamental to take into account the influence of dissipative mechanisms, such as the collisional interaction, to prevent the magnetic energy density of the ion-cyclotron modes from becoming negative at frequencies larger than the Hall frequency in the case of weakly ionized plasmas or the cyclotron frequency in fully ionized plasmas. In addition, the total energy density would also become negative for $\omega > 2 \Omega_{\rm{H}}$ and $\omega > 2 \Omega_{\rm{i}}$ for weakly and fully ionized plasmas, respectively, if collisions between the different components of the plasma are neglected. Regarding these results, it is worth to mention that at such high frequencies it may be more appropriate to use a three-fluid model \citep[see, e.g.,][]{Watanabe1961CaJPh..39.1044W,Tanenbaum1962PhFl....5.1226T,Pinto2008A&A...484...17P}, in which ions and electrons are treated separately, instead of resorting to the two-fluid model considered here, so the effect of the inertia of electrons is not neglected. Therefore, it would be interesting to check in the future the validity of these findings with a more accurate plasma model, but also by studying the full temporal evolution of the perturbations instead of just focusing on the properties of the normal modes.

    As concluding remarks, we discuss in the following lines some possible improvements to the present work. We remind that this is the first installment of a series devoted to the study of the properties of magneto-hydrodynamic waves propagating in weakly ionized plasmas. Here, we have focused only on small-amplitude incompressible perturbations, which correspond to the low frequency Alfvén waves, and the high-frequency ion-cyclotron and whistler waves. Therefore, the study of compressible perturbations, that is, magneto-acoustic waves, is left for a forthcoming paper. Compressibility adds new modes for wave propagation \citep[see, e.g.,][]{Lighthill1960RSPTA.252..397L,Goedbloed2004prma.book.....G,Zaqarashvili2011A&A...529A..82Z,Mouschovias2011MNRAS.415.1751M,Soler2013ApJS..209...16S} where the pressure forces may play a relevant role, so it is also of great interest to analyze their polarization relations and the distribution of their wave energy using a two-fluid model. In addition, it has been shown by \citet{Waters2013EP&S...65..385W}, \citet{Cally2015ApJ...814..106C}, \citet{GonzalezMorales2019ApJ...870...94G}, and \citet{Raboonik2019SoPh..294..147R,Raboonik2021MNRAS.507.2671R} that Hall's current produces a coupling between magneto-acoustic and Alfvén waves in weakly ionized plasmas. The efficiency of this coupling greatly depends on the relation between the frequency of the wave, the cyclotron frequency and the ionization degree of the plasma. Thus, it would be expected that the Hall contribution to the Poynting flux would no longer be $\bm{\Pi}_{\rm{Hall}} = 0$ but that it would have an increasing influence in the propagation of high-frequency magneto-acoustic waves.
    
    The present analytical investigation has only considered the linear regime of the plasma dynamics, so it cannot describe various relevant non-linear processes that take place in partially ionized plasmas, such as the heating due to the charge-neutral collisions \citep[see, e.g.,][]{Piddington1956MNRAS.116..314P,Osterbrock1961ApJ...134..347O,Leake2005A&A...442.1091L,Song2011JGRA..116.9104S,Arber2016ApJ...817...94A,Soler2016A&A...592A..28S,Khomenko2019ApJ...883..179K}, and the influence of the collisions on the evolution of shocks \citep{Hillier2016A&A...591A.112H,Popescu2019A&A...627A..25P,Popescu2019A&A...630A..79P,Snow2021A&A...645A..81S} or on the ponderomotive force that couples Alfvén and slow waves \citep{MartinezGomez2018ApJ...856...16M,Ballester2020A&A...641A..48B,Ballester2024RSPTA.38230222B}. These processes are more appropriately addressed by means of numerical simulations, which will be also used in the future to get a better understanding of the evolution of the energy of the waves and the relations between its components.
    
    Furthermore, here we have studied the role played by Hall's current on Alfvénic waves. However, we have neglected other non-ideal effects, such as resistivity (Ohmic diffusion), which has been shown to have a strong impact in the damping of the whistler modes at high frequencies \citep[see, e.g.,][]{MartinezGomez2017ApJ...837...80M}, or heat conduction and viscosity, which have been found to be more efficient than charge-neutral collisions in damping the slow magneto-acoustic waves in the solar chromosphere and prominences \citep{Forteza2008A&A...492..223F,Soler2010A&A...512A..28S,Barcelo2011A&A...525A..60B,Soler2015ApJ...810..146S}. These effects need to be taken into consideration altogether to get a more complete description of the dynamics of partially ionized plasmas. 
    
\begin{acknowledgements}
    DM acknowledges support from the Spanish Ministry of Science and Innovation through the grant CEX2019-000920-S of the Severo Ochoa Program. DM is also grateful to Elena Khomenko and Roberto Soler for their suggestions and discussions on the present results and to the anonymous referee for the useful corrections.
\end{acknowledgements}

\bibliographystyle{aasjournal}
\bibliography{waves_weakly}

\appendix
\section{Wavelengths and damping lengths} \label{sec:wavelengths}
\begin{figure} [h]
        \centering
        \includegraphics[width=\hsize]{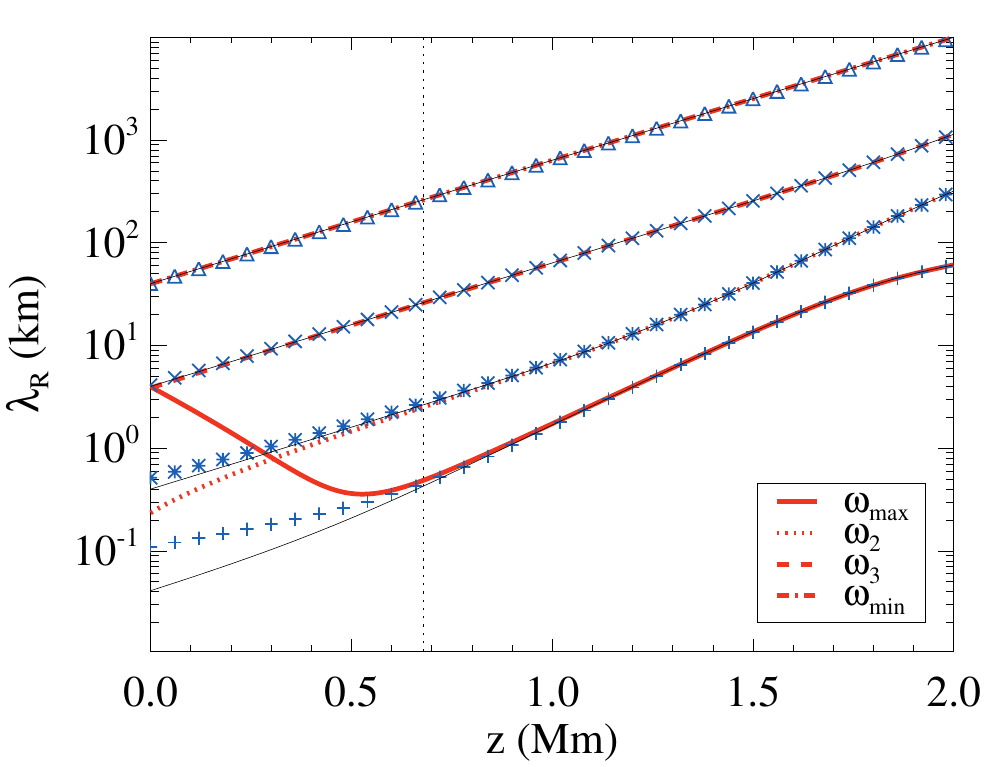} \\
        \includegraphics[width=\hsize]{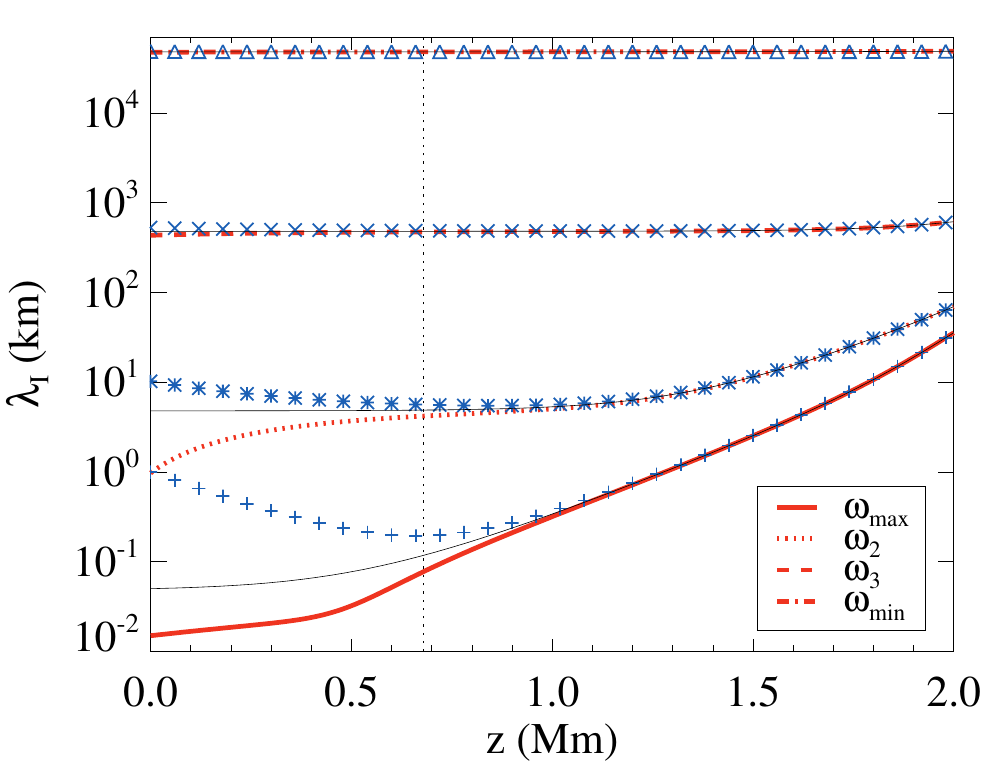}
        \caption{Wavelength $\lambda_{\rm{R}}$ (top panel) and damping length $\lambda_{\rm{I}}$ (bottom panel) as functions of height for the Alfvén (black thin lines), ion-cyclotron (red lines) and whistler (blue symbols) modes. The vertical dotted line marks the height where $\omega_{\rm{max}} = \Omega_{\rm{H}}$.}
        \label{fig:lambdas_hall}
    \end{figure}

    For the sake of completeness, we show here the solutions to Eq. (\ref{eq:kz2_hall}) in terms of the wavelengths and damping lengths of the waves, which are given by
    \begin{equation} \label{eq:lambda}
        \lambda_{\rm{R}} = \frac{2 \pi}{k_{\rm{R}}} \quad \text{and} \quad \lambda_{\rm{I}} = \frac1{k_{\rm{I}}},
    \end{equation}
    respectively. The results as functions of height are displayed in Fig. \ref{fig:lambdas_hall}, including the solutions for the linearly polarized Alfvén waves (black thin lines), and the circularly polarized ion-cyclotron and whistler modes (represented by the red lines and the blue symbols, respectively). We see in the top panel that the wavelength always increases with height except for the case of the ion-cylotron mode with $\omega_{\rm{max}}$ at the bottom layers of the atmosphere, which has a minimum wavelength around $z \approx 0.5 \ \rm{Mm}$. For the waves with the lowest frequencies ($\omega_{\rm{3}}$ and $\omega_{\rm{min}}$), which have periods of $10$ and $100$ seconds, the wavelength becomes of the order of $10^{3} \ \rm{km}$ (or larger) at the top layers, which is comparable with the size of the model atmosphere we have considered. Therefore, we expect that at such heights these waves are strongly affected by the gravitational stratification, as it is discussed in Appendix \ref{sec:strat}. At the bottom layers we see that the wavelength of the Alfvén modes is inversely proportional to the wave frequency and it is always smaller than the wavelength of the whistler modes. In this region of the atmosphere, the modes corresponding to the largest frequencies (and periods of $0.1$ and $1$ seconds) have wavelengths shorter than $1 \ \rm{km}$.
    
    The bottom panel of Fig. \ref{fig:lambdas_hall} shows that for the lowest frequency ($\omega_{\rm{min}} = 0.02 \pi \ \rm{rad \ s^{-1}}$) the damping lengths are much larger than $10^{4} \ \rm{km}$, meaning that the corresponding waves will cross the atmosphere without any noticeable decrease in their amplitude. Then, the damping length has a strong dependence on the inverse of the wave frequency, reaching values of the order of $10 \ \rm{m}$ for the highest frequency waves at the bottom layers of the atmosphere. The damping lengths of the ion-cyclotron modes are always smaller than those of the Alfvén waves, which are in turn smaller that the damping length of the whistler modes. This shows again that in the regime of high frequencies, the effect of charge-neutral collisions on the propagation of waves strongly depends on their polarization state, with the right-handed circularly polarized modes being able to propagate across larger distances than the other two kinds of waves.

\section{On the effect of gravitational stratification} \label{sec:strat}
\begin{figure} [h]
        \includegraphics[width=\hsize]{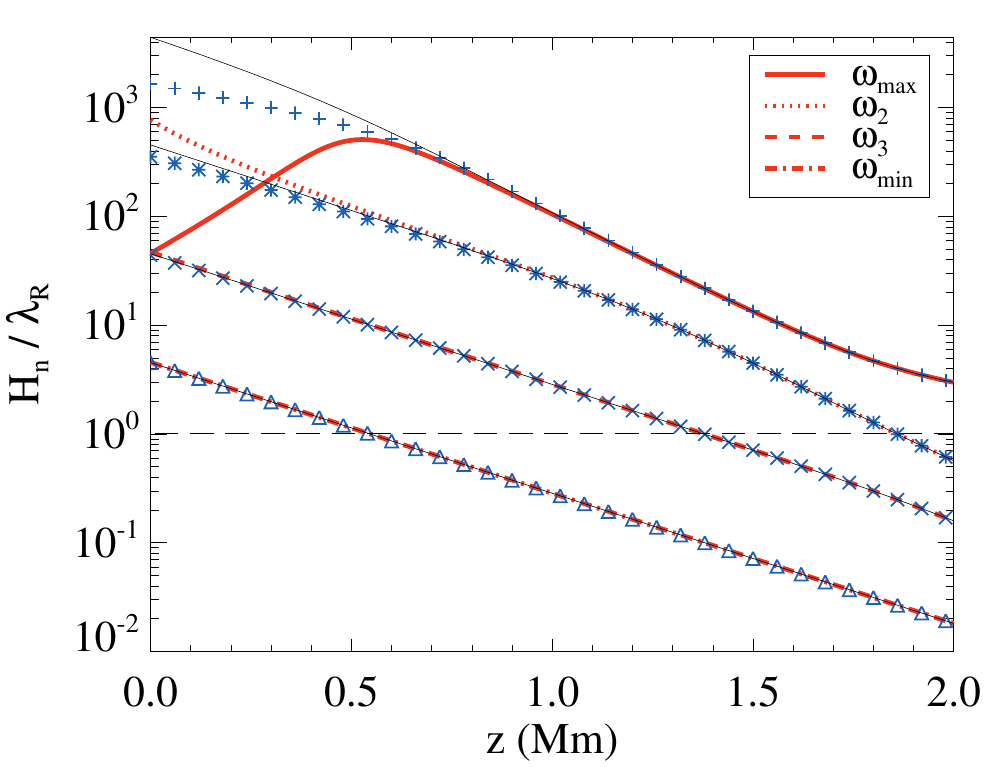}
        \caption{Comparison of the gravitational scale height, $H_{\rm{n}}$, and the wavelengths, $\lambda_{\rm{R}}$ as functions of height. The horizontal dashed line represents the value $H_{\rm{n}}/\lambda_{\rm{R}} = 1$. Same colors, lines, and symbols styles as in Fig. \ref{fig:lambdas_hall} have been used.}
        \label{fig:hhkr_hall}
    \end{figure}

    The results analyzed in the main text have been obtained by neglecting the influence of gravitational stratification on the wavelengths and damping lengths of the perturbations. As mentioned in Section \ref{sec:linear}, this has been justified on the assumption that the wavelengths resulting from the dispersion relation are much shorter than the vertical scale heights related to gravity. To check the validity of this assumption, we present in Fig. \ref{fig:hhkr_hall} the ratio $H_{\rm{n}} / \lambda_{\rm{R}}$ as a function of height. We have chosen the vertical scale height of the neutral fluid as the value of reference for this comparison because it is smaller than $H_{\rm{c}}$, so it is more restrictive, and also because we are analysing a weakly ionized atmosphere, so it is expected that (when the collisional coupling is strong) the physical conditions of the neutral fluid would have a larger impact on the hydrodynamics than those of the charged fluid.

    We see in Fig. \ref{fig:hhkr_hall} that at the bottom of the model atmosphere all the modes have a ratio $H_{\rm{n}} / \lambda_{\rm{R}} > 1$, so it is not expected that the gravitational stratification has any significant effect on the wavelength and damping rates of the perturbations. The ratio tends to increase with the frequency of the waves (except for the case of the the ion-cyclotron mode, which shows a smaller value for $\omega_{\rm{max}} = 20 \pi \ \rm{rad \ s^{-1}}$ than for $\omega_{\rm{2}} = 2 \pi \ \rm{rad \ s^{-1}}$). Hence, the assumption of neglecting the vertical stratification is more valid for the case of high-frequency waves. However, as we move towards the upper layers of the atmosphere we find that the ratio becomes $H_{\rm{n}} / \lambda_{\rm{R}} < 1$ (except for $\omega_{\rm{max}}$), specially for the longer periods, which show values of $H_{\rm{n}} / \lambda_{\rm{R}} \ll 1$. Thus, the assumption of neglecting the vertical stratification breaks for the low frequency waves and the analysis previously performed has to be taken with some caution, although it serves as a good first approximation to the properties of these waves. For the high-frequency waves the assumption can be safely applied at almost every height of the model atmosphere.

    Another important effect of the gravitational stratification that is not captured by our analysis is the growth of the velocity amplitude as the waves propagate from the bottom denser layers of the atmosphere to the top lighter layers \citep[see, e.g.,][]{Zaqarashvili2013A&A...549A.113Z}. Depending on the initial amplitude of the perturbation, this growth can lead to non-linear effects such as the development of shocks \citep{Montgomery1959PhRvL...2...36M,Cohen1974PhFl...17.2215C} or the generation of perturbations in density and pressure \citep{Hollweg1971JGR....76.5155H,Rankin1994JGR....9921291R}, which cannot be described with the linear method used here. Numerical simulations would have to be used to properly address the non-linear regime of waves propagating in a two-fluid stratified atmosphere. This has been done, for instance, by \citet{Popescu2019A&A...627A..25P,Popescu2019A&A...630A..79P}, who found that the collisional damping may reduce the formation of shocks and smooth the steep wave fronts.

\end{document}